\documentclass[12pt]{iopart}
\pdfoutput=1

\usepackage{iopams}
\usepackage{graphicx}
\usepackage{subfigure}
\usepackage[numbers,square,sort&compress]{natbib}
\usepackage{esint}

\usepackage[colorlinks=true,linkcolor=blue,urlcolor=blue,citecolor=blue,bookmarks]{hyperref}
\usepackage{url}

\usepackage[utf8]{inputenc}
\usepackage{ae}

\newcommand{\eg}{e.\,g.}
\newcommand{\ie}{i.\,e.}

\begin{document}

\title[Anisotropy in finite continuum percolation]{Anisotropy in finite continuum percolation:\\ Threshold estimation by Minkowski functionals}

\author{Michael A Klatt$^{1,2}$, Gerd E Schr\"oder-Turk$^3$ and Klaus Mecke$^2$}

\address{$^1$ Karlsruhe Institute of Technology (KIT), Institute of Stochastics, Englerstr. 2, 76131 Karlsruhe, Germany}
\address{$^2$ Institut f\"ur Theoretische Physik, Universit\"at Erlangen-N\"urnberg, Staudtstr. 7, 91058 Erlangen, Germany}
\address{$^3$ School of Engineering \& IT, Murdoch University, 90 South Street, Murdoch, WA 6150, Australia}
\ead{michael.klatt@kit.edu}

\begin{abstract}
We examine the interplay between anisotropy and percolation, \ie, the spontaneous formation of a system spanning cluster in an anisotropic model. 
We simulate an extension of a benchmark model of continuum percolation, the Boolean model, which is formed by overlapping grains.
Here we introduce an orientation bias of the grains that controls the degree of anisotropy of the generated patterns.
We analyze in the Euclidean plane the percolation thresholds above which percolating clusters in $x$- and in $y$-direction emerge.
Only in finite systems, distinct differences between effective percolation thresholds for different directions appear.
If extrapolated to infinite system sizes, these differences vanish independent of the details of the model.
In the infinite system, the uniqueness of the percolating cluster guarantees a unique percolation threshold.
While percolation is isotropic even for anisotropic processes, the value of the percolation threshold depends on the model parameters, which we explore by simulating a score of models with varying degree of anisotropy.
To which precision can we predict the percolation threshold without simulations?
We discuss analytic formulas for approximations (based on the excluded area or the Euler characteristic) and compare them to our simulation results.
Empirical parameters from similar systems allow for accurate predictions of the percolation thresholds (with deviations of $<5\%$ in our examples),
but even without any empirical parameters, the explicit approximations from integral geometry provide, at least for the systems studied here, lower bounds that capture well the qualitative dependence of the percolation threshold on the system parameters (with deviations of $5\%$--$30\%$).
As an outlook, we suggest further candidates for explicit and geometric approximations based on second moments of the so-called Minkowski functionals.
\end{abstract}

\vspace{2pc}

\submitto{\JSTAT}

\maketitle

Percolation is a geometric phase transition in which formerly local clusters spontaneous form a system spanning cluster. 
Dating back to the work by Flory~\cite{Flory1941} and Stockmayer~\cite{Stockmayer1944} and later by Broadbent and Hammersley~\cite{BroadbentHammersley1957},
percolation theory describes in general the critical behavior of connected clusters, either on lattices or in the continuum~\cite{Stauffer:1985, Grimmett2013}.
It is a general concept that can be applied to a multitude of very different physical systems and complex networks~\cite{Sahimi2013}.

In this paper, we are interested in the interplay of topology and geometry, more precisely, of percolation and anisotropy.
We investigate continuum percolation, that is, percolation on a continuous space, in statistically homogeneous (also called stationary) two-phase random media.
More specifically, we study a benchmark model of continuum percolation~\cite{MeesterRoy1996}, the so-called Boolean model.
It consists of overlapping grains that are distributed randomly in space.
The intensity~$\gamma$, which is also called number density, gives the expectation for the number of grain centers per unit area.
It is one of the parameters of the system.

The overlapping grains form clusters of connected components.
Starting from a dilute arrangement of grains, the mean size of these clusters grows with increasing area fraction of the grain phase, called the occupied area fraction.
Below the critical area fraction, the so-called percolation threshold, these clusters are almost surely (\ie, with probability one) bounded.
At occupied volume fractions above the percolation threshold, almost surely exactly one unbounded cluster emerges that spans the whole system.
It is called the percolating cluster.
Percolation is a geometric phase transition with universal scaling of geometrical properties, for example, the size of the percolating cluster.

Anisotropy appears in many man-made or natural materials, from geological formations~\cite{Postma1955} over fiber composites and laminates~\cite{christensen_mechanics_2005} to porous media~\cite{dullien_porous_1992}.
Moreover, anisotropy manifests itself in different forms.
A heterogeneous medium might be isotropic w.r.t. the distribution of the area but anisotropic w.r.t. of the orientation of the interface, or vice versa~\cite{SchroederTurketal2010AdvMater}.
Anisotropic two-phase random media can be modeled by the Boolean model with an orientation bias of the grains~\cite{SchroederTurketal2010AdvMater, HoerrmannEtAl2014}.

In these systems, percolation exhibits an interesting qualitatively different behavior in finite samples or in the idealized infinite limit.
Although there is no phase transition in a finite system, an effective percolation threshold can be defined for the ensemble.
As expected, it is anisotropic.
In other words, the system is more likely to percolate in the preferred direction than in the perpendicular direction.

However, this difference vanishes in the thermodynamic limit.
Even the most anisotropic system simultaneously percolates in all directions.
The percolation threshold is isotropic, see also~\cite{Balberg:1983, BalbergBinenbaumWagner1984}, which we here link to the uniqueness of the percolating cluster.
There have already been numerous studies of anisotropic continuum percolation, some using analytic calculations~\cite[\eg,][]{DroryEtAl1991, chatterjee_percolation_2014, chatterjee_tunneling_2015} others are based on Monte Carlo simulations~\cite[\eg,][]{Balberg:1983, kale_effect_2016} or experiments~\cite[\eg,][]{BalbergEtAl1983Experiment}.
There are also detailed studies of anisotropic lattice percolation~\cite{RednerStanley1979, Ikeda1979, Masihi:2006}.

While the isotropy of the percolation threshold is independent of the details of the model, the value of the percolation threshold is a nonuniversal constant, that is, it depends on the details of the system and specifically on its anisotropy.
For Boolean models an increasing degree of anisotropy typically also increases the percolation threshold of the grain phase~\cite{BalbergBinenbaumWagner1984, Chatterjee2015, kale_effect_2016}.
Because of the complexity of the geometric problem, analytic results for the percolation thresholds of Boolean models are hard to derive.
Precise numerical estimates require complex and time-consuming calculations.
Explicit formulas for estimates of percolation thresholds help to adjust models to experimental systems and predict their physical properties.
We determine, from simulations, the percolation thresholds for a score of models with varying degree of anisotropy.
How well can we predict the dependence of the critical intensity on the model parameters?

First, we compare the simulation results to the well-known excluded-area approximation, which uses an empirical parameter~\cite{Balberg:1984, BalbergBinenbaumWagner1984}.
Then, we examine a purely geometric lower bound without any empirical parameter, which is based on a topology index from integral geometry, the Euler characteristic~\cite{MeckeWagner1991}.
The simulation results suggest that the lower bound qualitatively captures well how the critical intensity depends on the model parameters.
We also suggest new candidates of purely geometric approximations based on further integral geometric shape descriptors.

In Section~\ref{sec:empirics}, we simulate anisotropic Boolean models (see Section~\ref{sec_anisotropic_Boolean_models}) and examine the behavior of continuum percolation both in finite samples and extrapolated to infinite systems.
For finite systems, we define the effective percolation threshold in Section~\ref{sec:percolation_connectivity} and extrapolate it to infinite system sizes in Section~\ref{sec_iso_perco}.
There we also provide several heuristic explanations of the empirical finding of an isotropic percolation threshold.
In Section~\ref{sec_perco_thresh_estimate}, we examine how the value of the percolation threshold changes with the anisotropy of the system.

In Section~\ref{sec:approx}, we study analytic approximations of the percolation thresholds based on the excluded area (Section~\ref{sec:excluded-area-approx}),
the Euler characteristic (Section~\ref{sec:euler}), or the second moments of Minkowski functionals (Section~\ref{sec_percolation_cov}).
In Section~\ref{sec_conclusion_and_outlook}, besides concluding remarks, we given an outlook to open problems and generalizations as well as to how our results suggest techniques for future research~\cite{Klatt2016}.


\section{Anisotropic ``percolation'' behavior in finite samples and isotropic thresholds in infinite systems}
\label{sec:empirics}

A percolating cluster in an infinite system is defined as an unbounded cluster; that is, a cluster which is not the subset of a compact observation window.
The percolation threshold in the intensity $\gamma_c$ is the intensity at which a percolating cluster emerges, more precisely, it is the infimum of all intensities $\gamma$ at which a percolating cluster appears (almost surely).
Alternatively, the percolation threshold $\phi_c$ in the occupied area fraction can be considered (if the mean area of a typical grain is strictly positive).
A phase transition occurs if the percolation threshold takes a non-trivial value $\gamma_c \in (0,\infty)$ (\eg, for Boolean models with a bounded grain size).
Above the percolation threshold, the probability that the cluster spans an observation window converges in the limit of an infinitely large window to unity.

\begin{figure}[t]
  \hspace*{0.16\textwidth}%
  \includegraphics[width=0.55\textwidth]{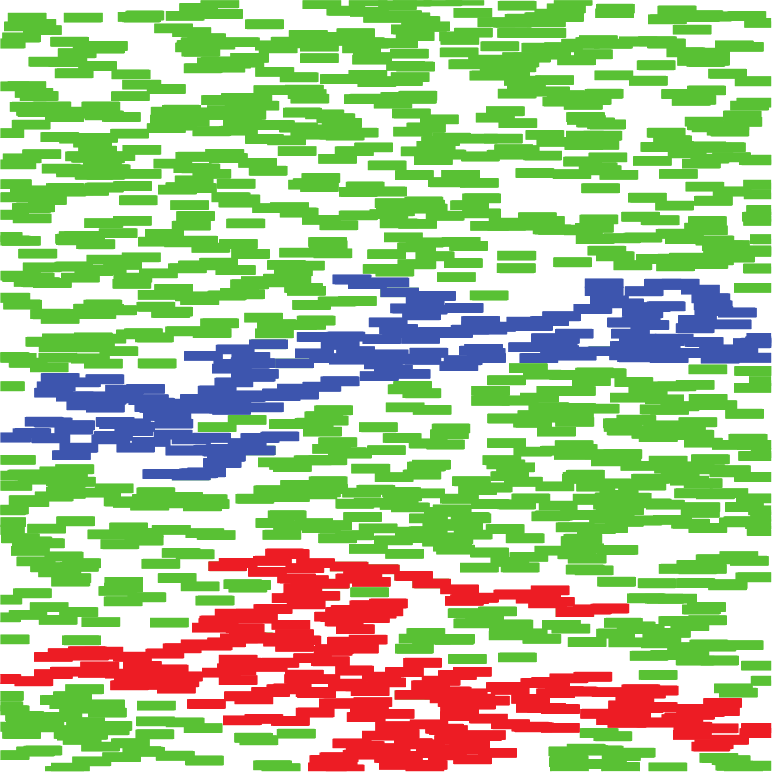}
  \caption{In a finite sample, a percolating cluster is defined as a cluster of particles that connects opposite sides of the sample.
           It spans the system in $x$- or in $y$-direction, respectively.
           The figure shows a Boolean model of fully aligned rectangles.
           The two percolating clusters in $x$-direction, which span the finite sample horizontally, are colored blue and red.
           There is no connected path spanning the system vertically.}
  \label{fig_perco_sample}
\end{figure}

In a finite sample, no infinite cluster can appear.
The percolating cluster is therefore defined as a cluster that connects opposite sides of the sample.
In other words, a cluster that percolates in $x$- or $y$-direction spans the system horizontally or vertically, respectively.
Figure~\ref{fig_perco_sample} shows a finite sample in which two clusters span the system in $x$- but none in $y$-direction.
Phase transitions occur only in the thermodynamic limit.
However, there is an apparent percolation threshold close to which the probability for the appearance of a percolating cluster steeply increases.

If the elongated particles have an orientation bias in $x$-direction, a finite anisotropic sample percolates more likely in $x$- than in $y$-direction.
In other words, the occupied volume fraction at which the probability that the system percolates in $x$-direction with a given probability is lower than the corresponding volume fraction in $y$-direction.
However, we show in this section that the difference vanishes in the limit of an infinite system size, no matter how anisotropic the system may be.
We relate this observation to the uniqueness of the percolating cluster.

\subsection{Anisotropic Boolean models}
\label{sec_anisotropic_Boolean_models}

The (isotropic) Boolean model is not only a benchmark model of continuum percolation but also a popular model of heterogeneous materials~\cite{Matheron:1975, Stoyan:2005}.
Examples include fractured materials or hydrating cement-based materials~\cite{GarbocziEtAl2011}, sedimentary rock~\cite{SchwartzEtAl1993, Martys:1994} and wood composites~\cite{WangShaler1998}.

\begin{figure}[t]
  \centering
  \hspace*{0.16\textwidth}%
  \subfigure[][]{\includegraphics[width=0.42\linewidth]{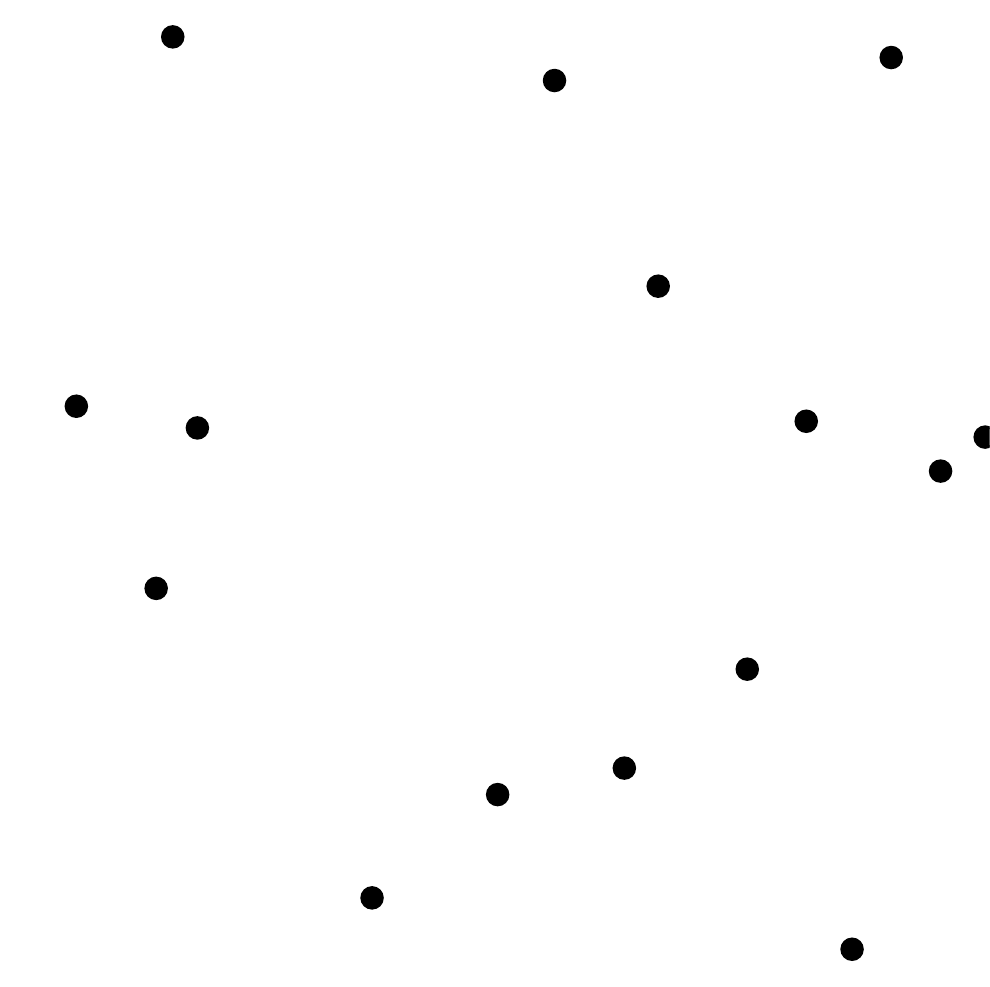}}%
  \subfigure[][]{\includegraphics[width=0.42\linewidth]{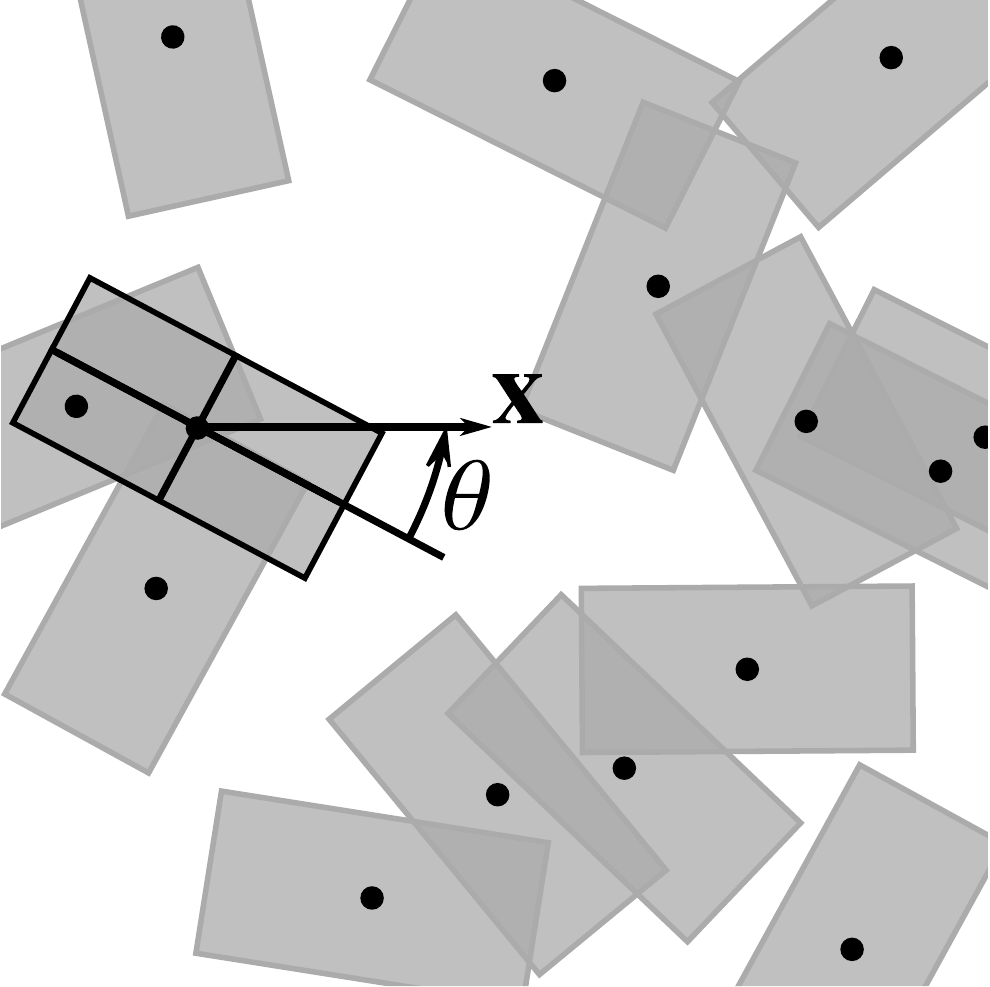}}%
  \caption{A Boolean model is formed by overlapping grains that are randomly placed in space distributed uniformly.
    (a) First, non-interacting points are placed in the plane.
    (b) Then, the points are decorated with grains, here rectangles with constant size and aspect ratio but a random orientation.
    The orientation of a grain is characterized by the angle $\theta$ between its main axis and the $x$-axis.}
  \label{fig_Boolean_model_def}
\end{figure}

Figure~\ref{fig_Boolean_model_def} visualizes the definition of the Boolean model.
First, points are randomly placed in space distributed uniformly.
They form a completely independent point process, the so-called Poisson point process.
The number of points in a finite observation window follows a Poisson distribution.
In the second step, each point is decorated with a grain.
The shape and orientation follows the so-called grain distribution.
We here only consider stationary processes, that is, statistically homogeneous models for which both intensity and orientation distribution of the grains are spatially constant.

\begin{figure}[t]
  \hspace*{0.16\textwidth}%
  \includegraphics[width=0.84\textwidth]{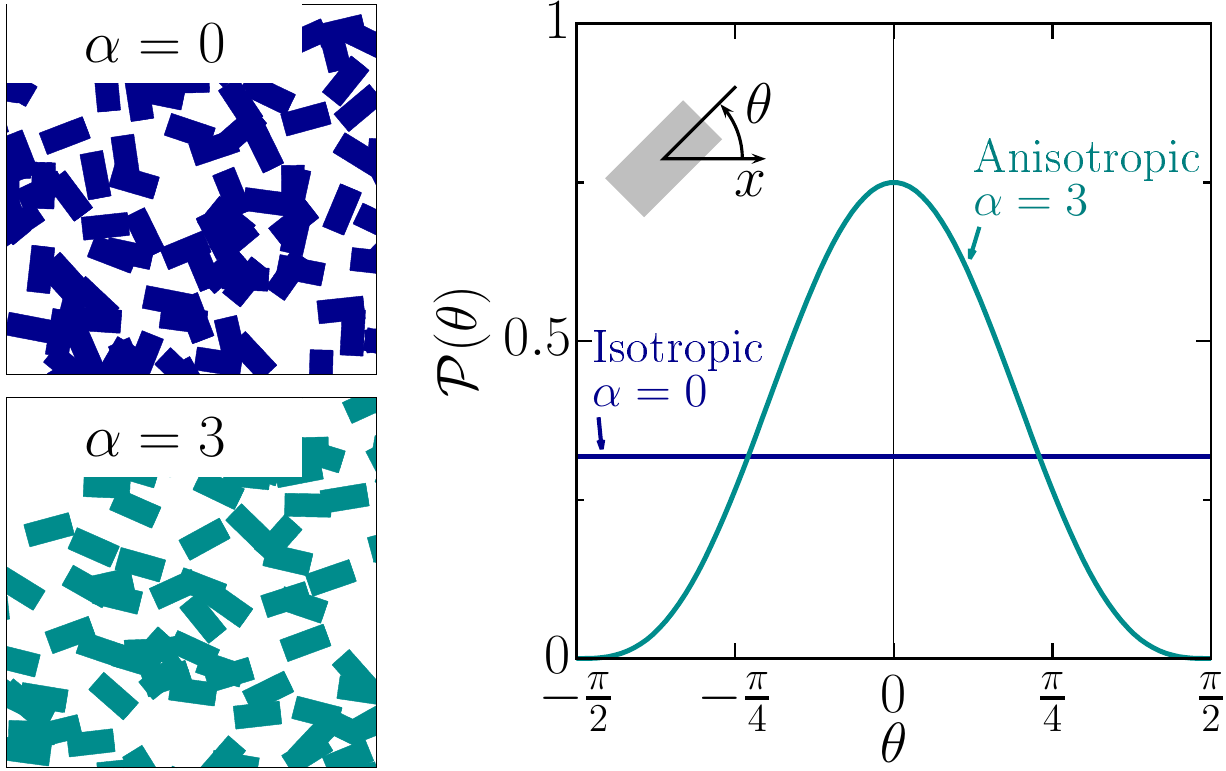}
  \caption{Orientation distributions of individual grains in Boolean models with anisotropy parameter $\alpha\in[0,\infty]$.
  The orientation of a single grain is parametrized by the angle $\theta$ between the main axis of the grain and the $x$-axis of the coordinate system.
  The probability density function $\mathcal{P}(\theta)\propto \cos^\alpha(\theta)$ of this angle determines the anisotropy of the Boolean model.
  Two samples are depicted on the left hand side of the figure.}
  \label{fig_mil_orientation_distribution}
\end{figure}

We use grains with a constant shape, namely, rectangles with a fixed size and aspect ratio.
To explore a range of different degrees of anisotropy in our simulations, we use a parametric orientation distribution from~\cite{HoerrmannEtAl2014}.
The anisotropy can be tuned by adjusting the orientation distribution of the grains.
The orientation of a grain is defined by the angle $\theta$ between its main axis and the $x$-axis.
Its probability density function is
\begin{equation}
  \mathcal{P}(\theta) := \frac{\Gamma(\frac{\alpha}{2}+1)}{2\sqrt{\pi}\,\Gamma(\frac{\alpha+1}{2})} \cdot \cos^{\alpha}(\theta),\quad \mathrm{for\ } \theta \in [-\frac{\pi}{2},\frac{\pi}{2}),
  \label{mil_orientation_distribution}
\end{equation}
see figure~\ref{fig_mil_orientation_distribution}.
The parameter $\alpha$ controls the degree anisotropy of the system:
$\alpha=0$ results in a uniform distribution and hence to an isotropic system;
the larger $\alpha$ the stronger the anisotropy;
we define that $\alpha=\infty$ corresponds to a full alignment with the $x$-axis.

The expected area fraction of the grain phase is called the occupied area fraction $\phi$.
It is independent of the orientation distribution and given by
\begin{equation}
  \phi = 1-\exp({-\gamma\,A})
  \label{eq:phi-gamma}
\end{equation}
where $A$ is the area of a single grain~\cite{SchneiderWeil2008}.
A dependency on the intensity always corresponds to an dependency on the occupied area fraction.


\subsection{Connectivity and effective percolation thresholds}
\label{sec:percolation_connectivity}

\begin{figure}[tb]
  \hspace*{0.16\textwidth}%
  \subfigure[][]{\includegraphics[height=5.75cm]{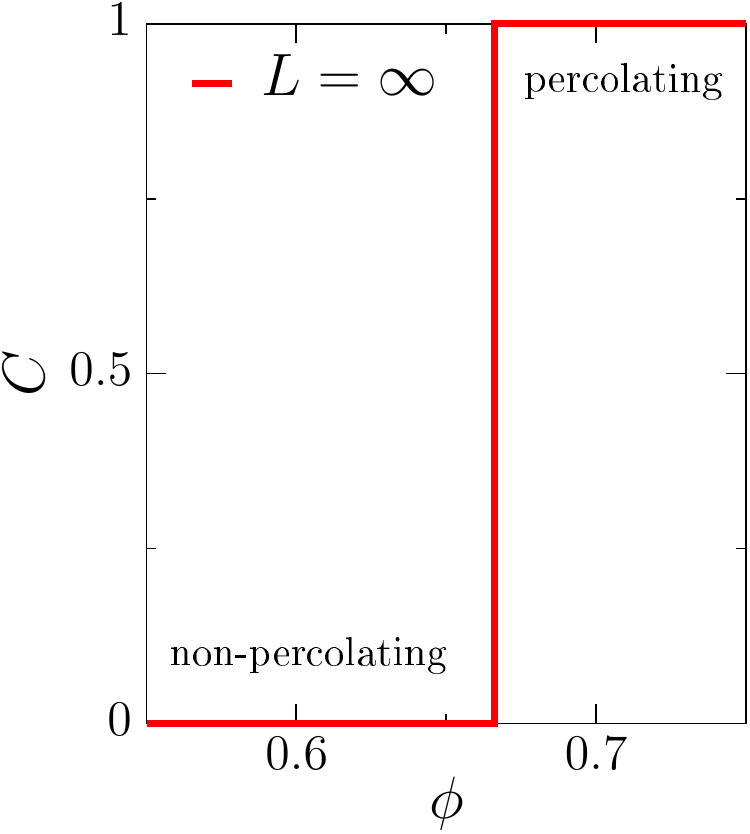}
  \label{fig_perco_geom_phase_trans}}
  \hfill
  \subfigure[][]{\includegraphics[height=5.75cm]{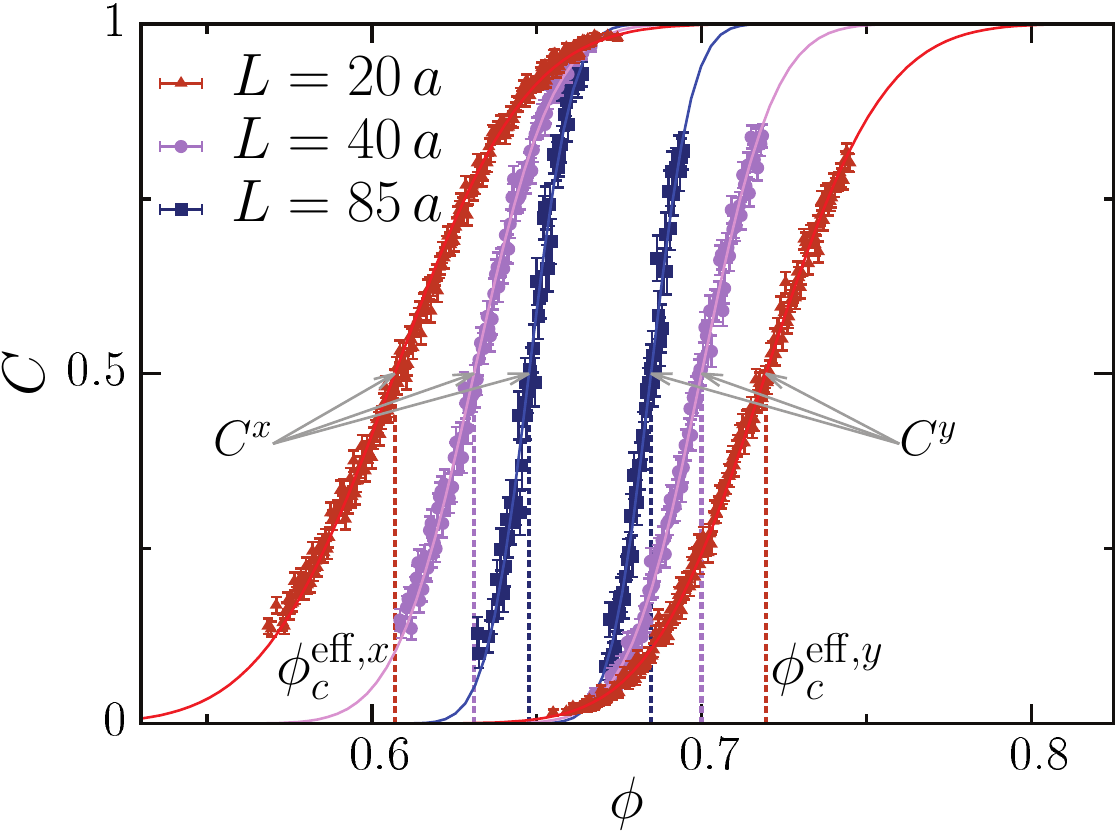}
  \label{fig_perco_finite_system_connectivity}}
  \caption{Connectivity $C$ as a function of the occupied area fraction $\phi$ for Boolean models: (a) for an infinite system size the connectivity is zero for $\phi < \phi_c$ and one for $\phi > \phi_c$;
    (b) for finite anisotropic systems the connectivity is a smooth function.
    It can take on values $0<C<1$.
    Most interestingly, it is different for percolation in $x$- or in $y$-direction.
    The plot shows numerical estimates of $C$ for a Boolean model of aligned rectangles with aspect ratio $1/4$ for different system sizes $L$ in units of the long side length of the rectangle.
    To each curve an error function is fitted according to (\ref{eq:erffit}).}
  \label{fig:connectivity}
\end{figure}

Percolation in an infinitely large system is a geometric phase transition.
Above the critical occupied area fraction $\phi_c$ there is a percolating cluster with probability one, and below this threshold there is almost surely no percolation of the grains.
The probability that there is a percolating cluster as a function of the occupied area fraction $\phi$ is called the connectivity $C$.
It is a step function that changes its functional value at $\phi_c$ from zero to unity, see figure~\ref{fig_perco_geom_phase_trans}.

We are interested in the behavior of finite systems.
In this case, there is no phase transition.
Depending on statistical fluctuations, a system spanning cluster may appear or it may not appear.
For all occupied area fractions $0 < \phi <1$ there is a finite probability that there is a percolating cluster.
The connectivity is expected to be a smooth and to take on values $0<C<1$.

The connectivity can be estimated from simulated Boolean models simply by the fraction of samples that contain a percolating cluster.
We use polygons for an exact representation of the Boolean model to avoid strong pixelation errors that would occur in pixelated Boolean models even for fine pixelations (because unconnected clusters can become connected or vice versa).
We simulate anisotropic Boolean models at different expected area fractions and for various sizes of the simulation box, and determine in each case the connectivity.
Figure~\ref{fig_perco_finite_system_connectivity} shows (for a Boolean model of aligned rectangles with aspect ratio $1/4$), as expected, a smooth increase of the connectivity for an increasing occupied area fraction $\phi$.

In figure~\ref{fig_perco_finite_system_connectivity}, we distinguish the connectivity in $x$- and $y$-direction in a Boolean model with fully horizontally aligned rectangles ($\alpha=\infty$).
The connectivity $C^x$ is the probability that there is a cluster that spans the system in $x$-direction,
and accordingly the connectivity $C^y$ considers clusters that span the system in $y$-direction.
More particles are needed for vertical than for horizontal percolation, and a fluctuation that spans the system horizontally is more likely than vertically.
The system is more likely to percolate horizontally than vertically (at least close to the percolation transition).
Even for relatively large system sizes, there are occupied volume fractions at which the system most likely percolates in $x$- but not in $y$-direction.

Since there is no phase transition in a finite system, the percolation threshold $\phi_c$ is not well-defined.
Instead an \textit{effective percolation threshold} $\phi_c^{\mathrm{eff}}$ must be defined.
A simple and intuitive definition is via
\begin{eqnarray}
  C(\phi_c^{\mathrm{eff}})=\frac{1}{2}\,. 
  \label{eq_perco_def_phieff}
\end{eqnarray}
A trivial estimate inverting the estimated connectivity as a function of the occupied area fraction would not be robust against statistical fluctuations.
A more robust estimate has been defined in \cite{Rintoul:1997}.
The derivative of the connectivity ${\mathrm{d}C}/{\mathrm{d}\phi}$ as a function of the occupied area fraction $\phi$ is the probability that the system starts to percolate at $\phi$.
Another possibility to define an effective percolation threshold is $\phi_c^{\mathrm{eff}}:=\int_0^{1} \phi({\mathrm{d}C}/{\mathrm{d}\phi})\rmd\phi$~\cite{Stauffer:1985}.
Because of the phase transition, the behavior of the system is close to the percolation threshold and for large systems dominated by the universal scaling.
If the derivative of the connectivity ${\mathrm{d}C}/{\mathrm{d}\phi}$ is approximated by a Gaussian distribution, the connectivity can be approximated by an error
function with a constant offset and prefactor:
\begin{eqnarray}
  {C}(\phi) &\approx
  \frac{1}{2}+\frac{1}{2}\,\mathrm{erf}\left(\frac{\phi-\phi_c^{\mathrm{eff}}}{\Delta}\right)\;.
  \label{eq:erffit}
\end{eqnarray}
Fitting this rescaled error-function to the connectivity as a function of $\phi$, where $\Delta$ and $\phi_c^{\mathrm{eff}}$ are the fit parameters, see figure~\ref{fig_perco_finite_system_connectivity}, provides a numerically robust estimate of the effective percolation threshold $\phi_c^{\mathrm{eff}}$.
In the limit of infinite system size, the error function converges to a step function and $\phi_c^{\mathrm{eff}}\rightarrow\phi_c$.

We adopt this procedure, but we use the intensity $\gamma$ as an input instead of the occupied area fraction $\phi$.
In a finite system, there are statistical fluctuations in $\phi$.
To avoid systematic errors, we replace the volume fractions in (\ref{eq:erffit}) by the corresponding intensities.

The definition of the effective percolation threshold via (\ref{eq:erffit}) has the advantage that it can be easily interpreted and implemented.
However, there are two major disadvantages.
First, unnecessary samples are simulated especially at high thresholds; they are computationally expensive to simulate because of the large number of grains.
Second, the precision is limited by unknown systematic errors.
Although the error function is a good approximation of the connectivity as a function of the occupied area fraction, high precision estimates of the connectivity reveal in finite systems statistically significant deviations from an error function.
The statistical errors in our simulation are of the order $\mathcal{O}(10^{-4})$, whereas systematic errors are of $\mathcal{O}(10^{-3})$.
There are more efficient algorithms using inhomogeneous Boolean models for more precise estimates of the percolation threshold of homogeneous models~\cite[\eg,][]{Quintanilla:1999, Quintanilla:2000, LorenzZiff2001}.
In future research, these algorithms could be generalized to anisotropic systems.
For our analysis of anisotropic continuum percolation, the procedure based on (\ref{eq:erffit}) is sufficient.

An error function is fitted to each curve in figure~\ref{fig_perco_finite_system_connectivity}.
The resulting estimates of the effective percolation thresholds are depicted by vertical dashed lines.
The effective percolation thresholds in $x$-direction are distinctly smaller than in $y$-direction for all simulated systems sizes $L$.
However, the differences in the effective percolation thresholds depend on the system sizes.
The differences for horizontal and vertical directions become smaller with increasing system size.

\subsection{Isotropic percolation threshold}
\label{sec_iso_perco}

Because percolation is a phase transition, the scaling of the effective percolation thresholds with the system size is universal for large system sizes.
The scaling exponent does not depend on the grain distribution, it is even the same for continuum percolation and lattice percolation; see \cite[\eg,][]{GawlinskiStanley1981, VicsekKertesz1981}:
\begin{eqnarray}
  \phi_c^{\mathrm{eff}}-\phi_c \propto L^{-1/\nu}
  \label{eq:scaling}
\end{eqnarray}
with the universal critical exponent $\nu=\frac{4}{3}$~\cite{Stauffer:1985}.

\begin{figure}[t]
  \hspace*{0.16\textwidth}%
  \includegraphics[width=0.479\textwidth]{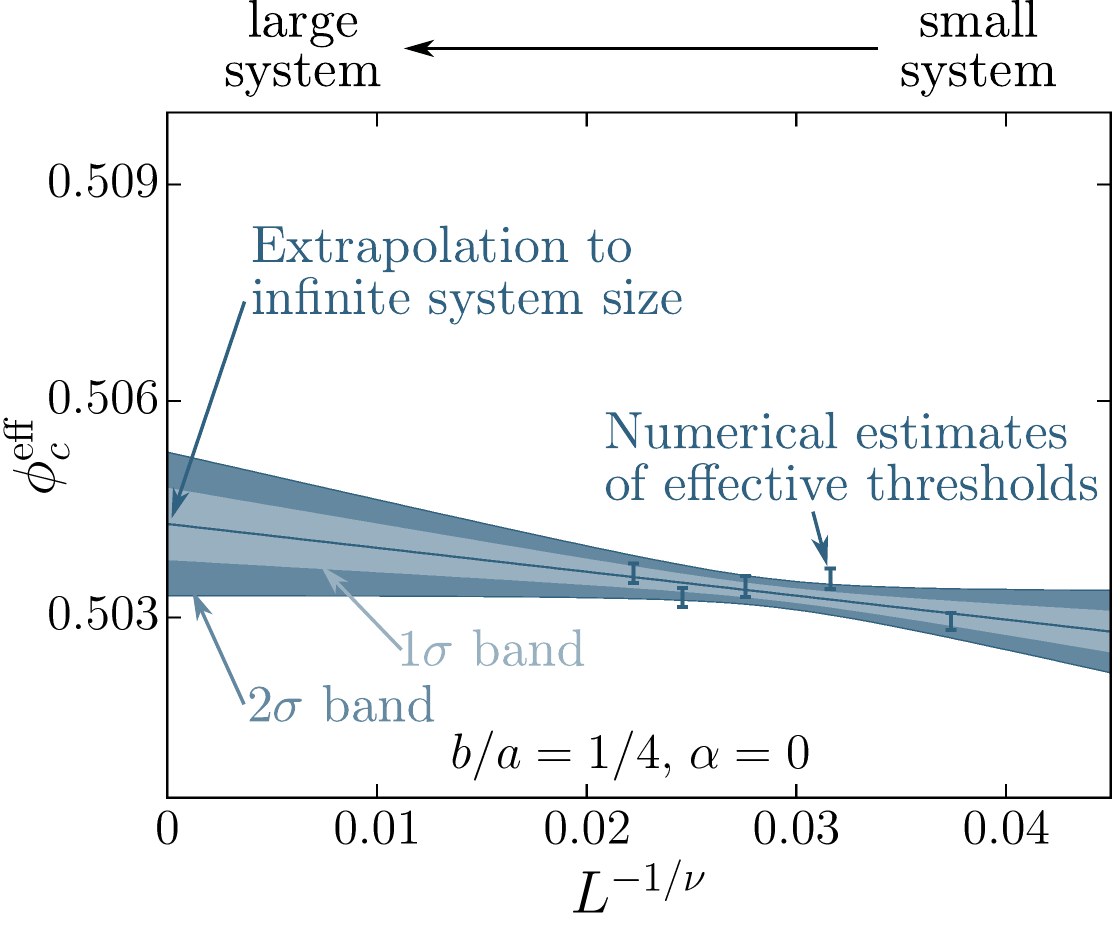}
  \caption{Finite size scaling of the effective percolation thresholds $\phi_c^{\mathrm{eff}}$, see (\ref{eq:scaling}), for an isotropic Boolean model ($\alpha=0$) with aspect ratio $b/a=\frac{1}{4}$.
  The unit of length for the linear system size $L$ is the long semiaxis of the rectangle.
  The color bands represent the bands of one and two standard deviations.}
  \label{fig_perco_scaling_isotropic}
\end{figure}

The percolation threshold $\phi_c$ can be estimated from the effective thresholds $\phi_c^{\mathrm{eff}}$ by an extrapolation from the finite system sizes $L$ to $L\rightarrow \infty$.
Figure~\ref{fig_perco_scaling_isotropic} shows for an isotropic Boolean model with rectangles with aspect ratio $b/a=\frac{1}{4}$ the numerical estimates of $\phi_c^{\mathrm{eff}}$, which are derived from fits of error functions to the connectivity (in $x$-direction).
It also plots the extrapolation to $\phi_c$ and the corresponding bands of one or two standard deviations.

\begin{figure}[p]
  \centering
  \subfigure{%
    \includegraphics[width=0.479\textwidth]{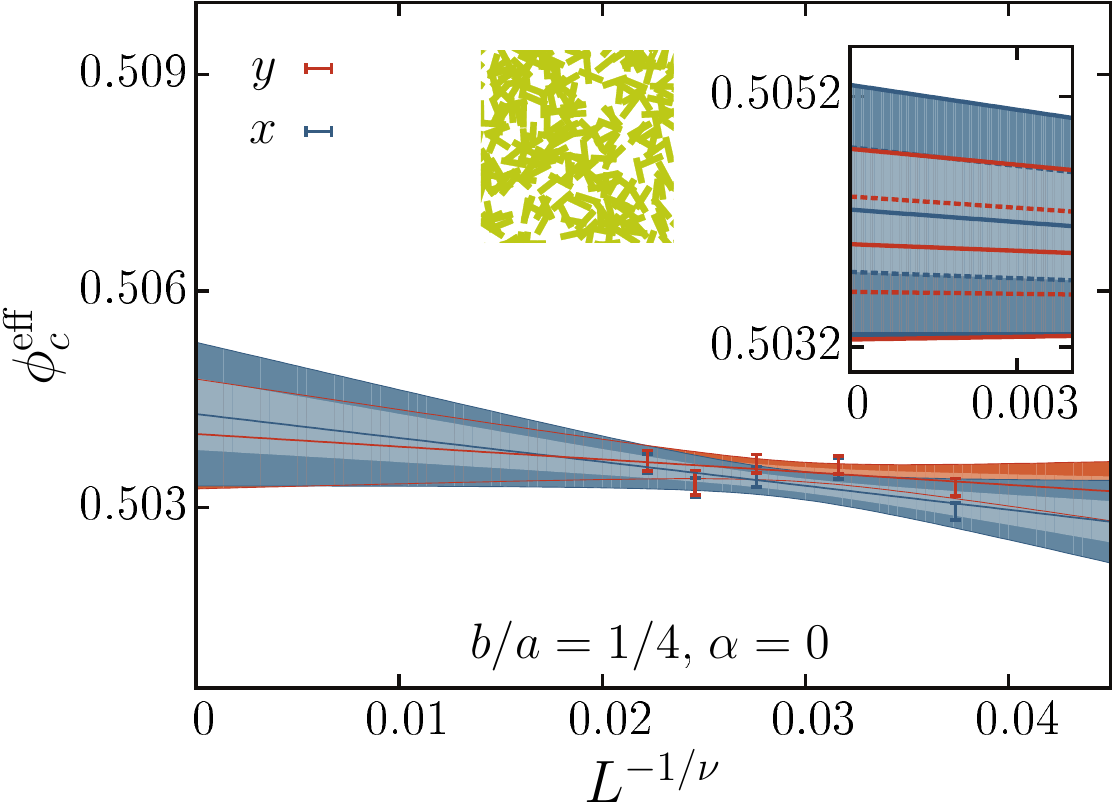}
  }%
  \hfill%
  \subfigure{%
    \includegraphics[width=0.479\textwidth]{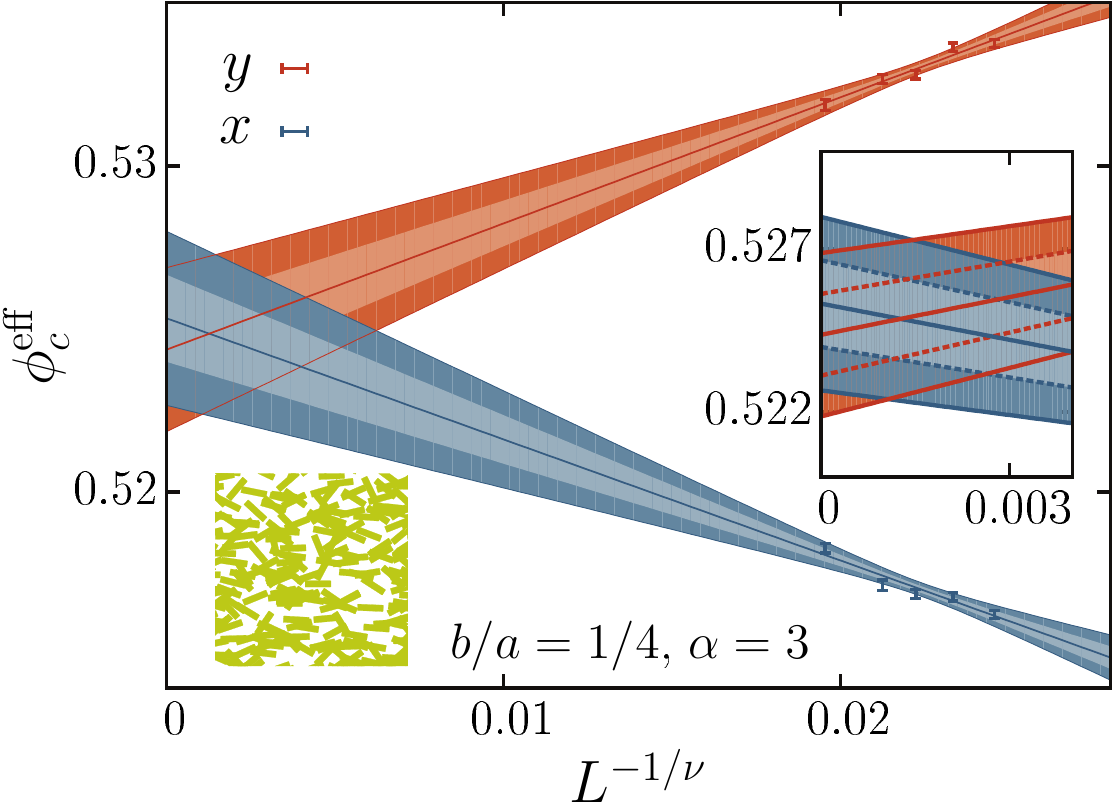}
  }\\
  \medskip
  \subfigure{%
    \includegraphics[width=0.479\textwidth]{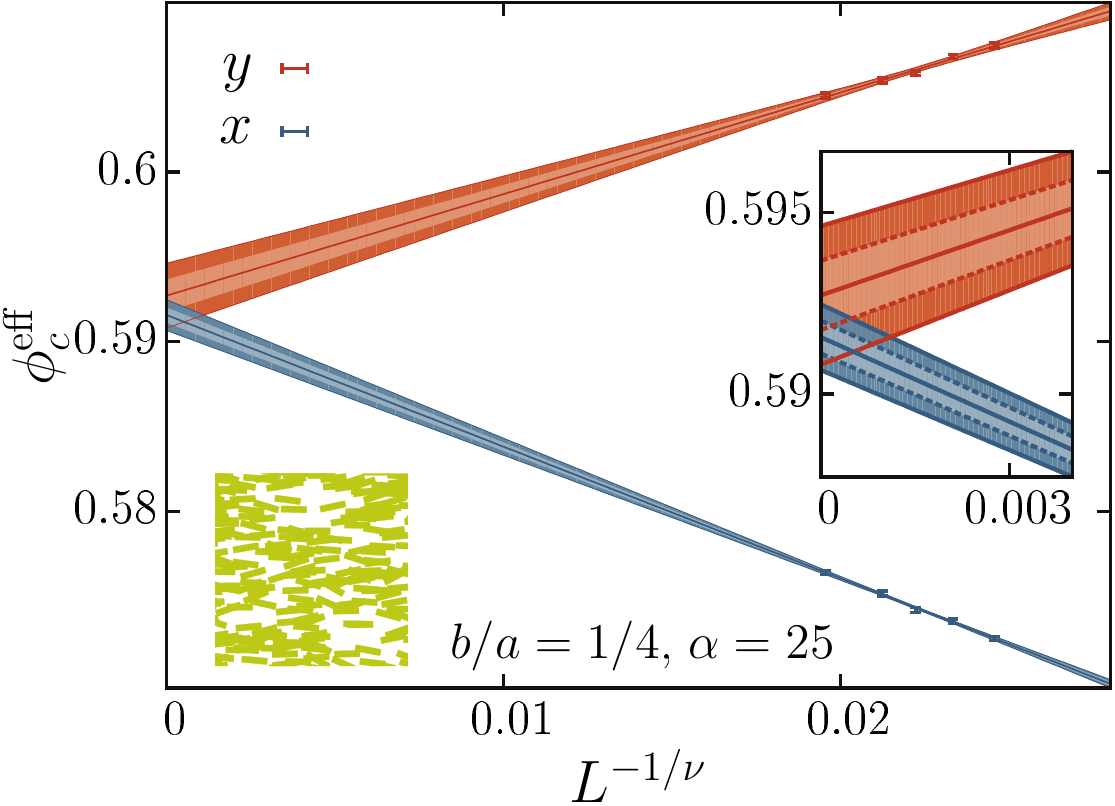}
  }%
  \hfill%
  \subfigure{%
    \includegraphics[width=0.479\textwidth]{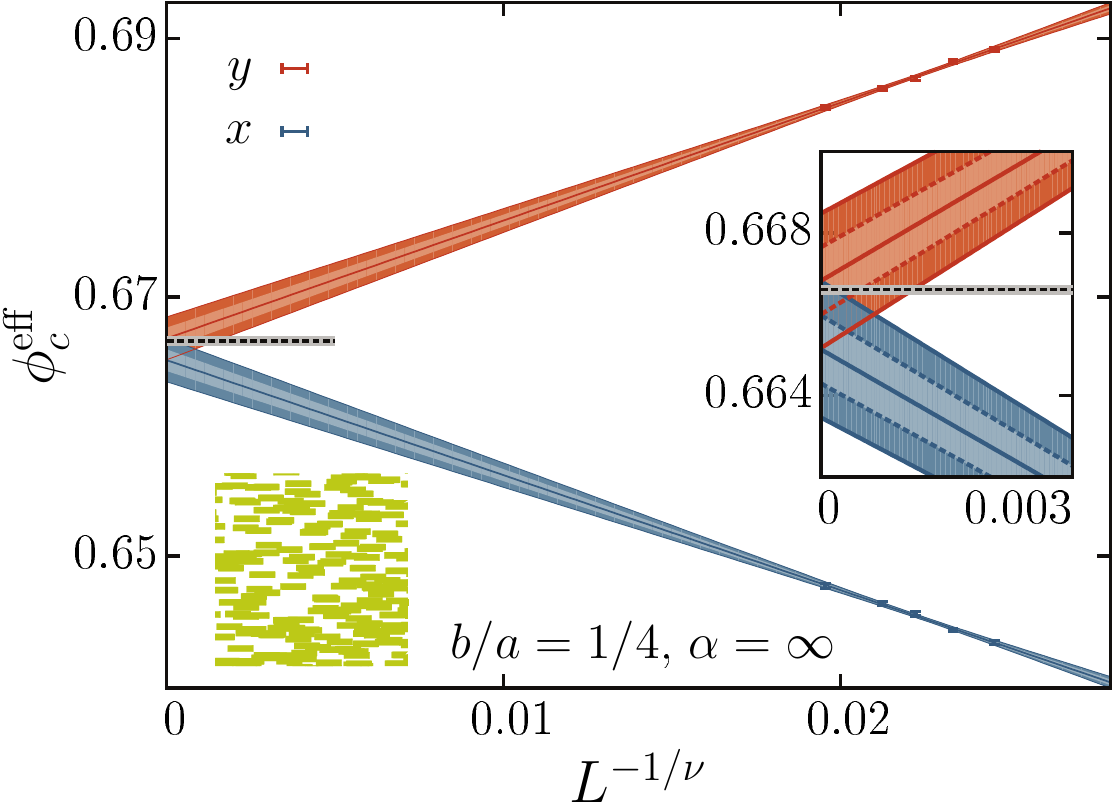}
  }\\
  \medskip
  \subfigure{%
    \includegraphics[width=0.479\textwidth]{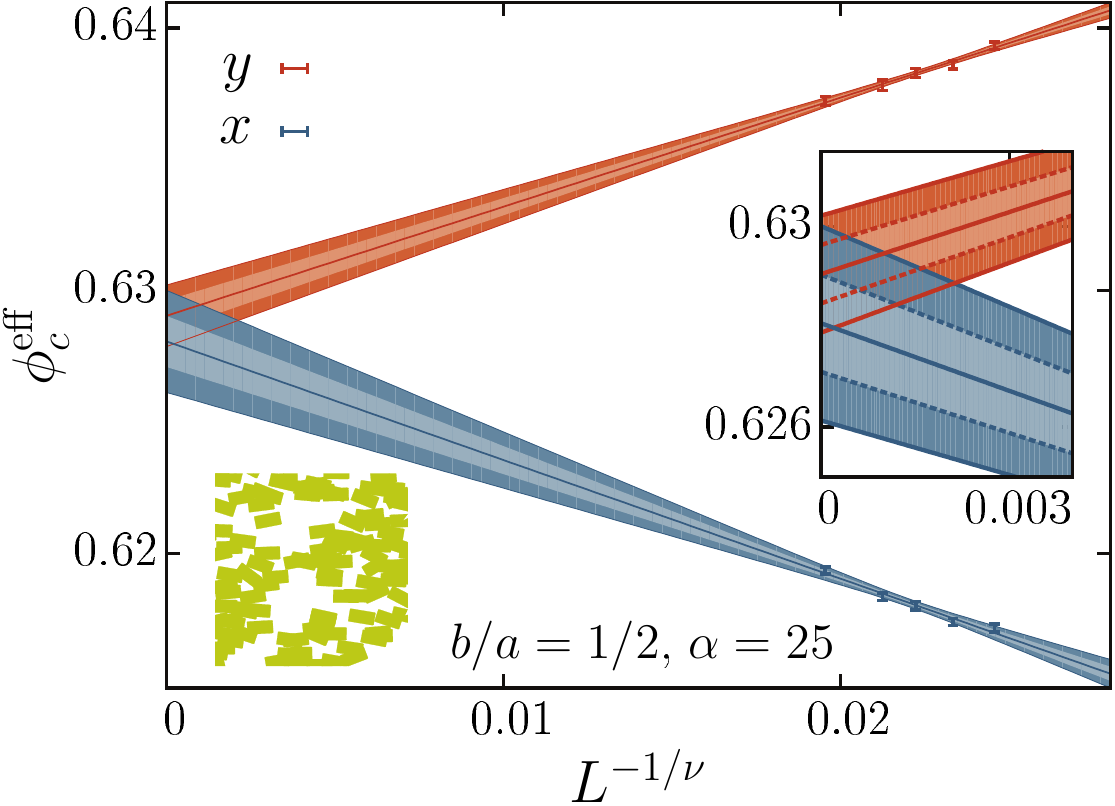}
  }%
  \hfill%
  \subfigure{%
    \includegraphics[width=0.479\textwidth]{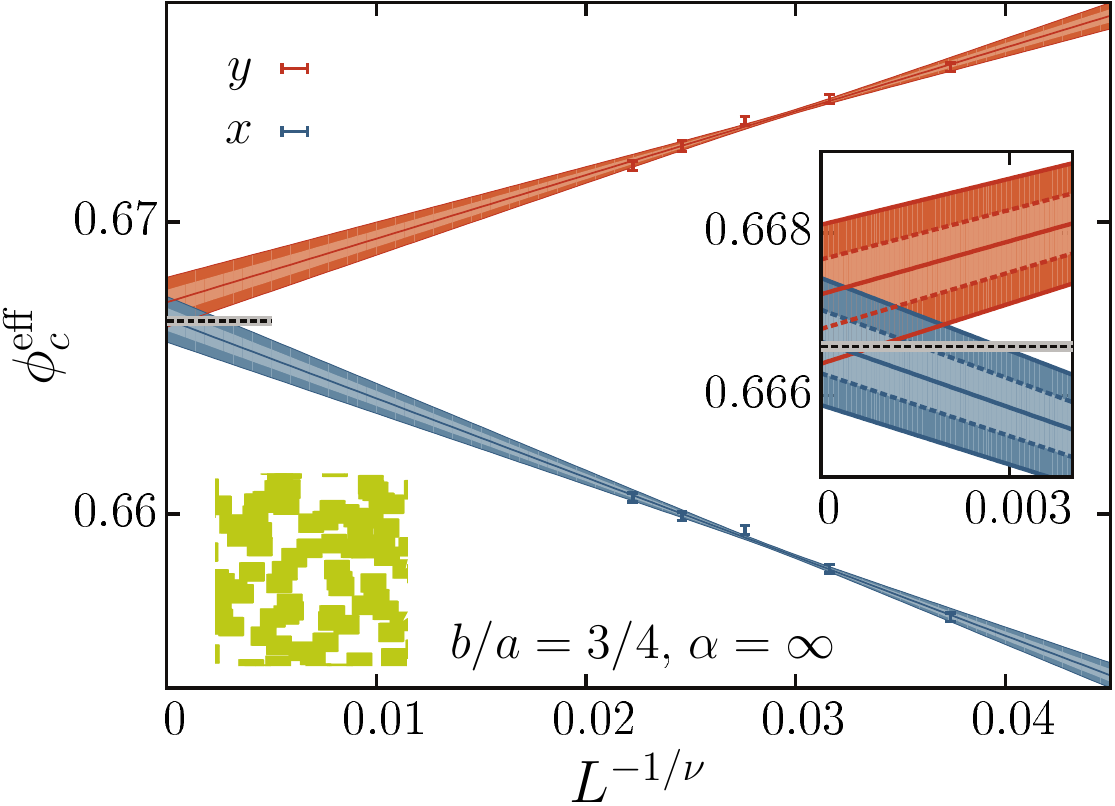}
  }
  \caption{Finite size scaling of the effective percolation thresholds that are extrapolated separately for the $x$- and the $y$-direction, see (\ref{eq:scaling_xy}).
          The unit of length for the linear system size $L$ is the long semiaxis of the rectangle.
          Each plot analyzes a different Boolean model consisting of rectangles with aspect ratio~$b/a$.
	  The anisotropy $\alpha$ varys from isotropy ($\alpha=0$) to full alignment ($\alpha=\infty$), see (\ref{mil_orientation_distribution}) and figure~\ref{fig_mil_orientation_distribution}.
	  In each plot, a sample of the corresponding Boolean model is depicted.
          Except for the isotropic system (top left), the effective percolation thresholds in $x$-direction are significantly below those in $y$-direction even for large systems.
          However, the extrapolated percolation thresholds of infinite systems are isotropic within statistical significance (the color bands represent the bands of one and two standard deviations).
          For aligned rectangles, the estimate and the error bar of the percolation threshold of aligned squares from~\cite{Baker:2002} are depicted by the dotted black line and gray band.}
  \label{fig_perco_scaling}
\end{figure}

For anisotropic models, we have to consider that the proportionality constant is nonuniversal.
It varies for the effective percolation thresholds $\phi_c^{\mathrm{eff},x}$ in $x$- and $\phi_c^{\mathrm{eff},y}$ in $y$-direction.
Therefore, we a priori distinguish also in the limit of infinite system size the percolation thresholds in $x$- or in $y$-direction.
In other words, we estimate the thresholds in $x$- or in $y$-direction separately:
\begin{eqnarray}
    \phi_c^{\mathrm{eff},x} &= \phi_c^x + m^x L^{-1/\nu}\ \mathrm{and}\\
    \phi_c^{\mathrm{eff},y} &= \phi_c^y + m^y L^{-1/\nu}.
  \label{eq:scaling_xy}
\end{eqnarray}
However, we will show that even for the most anisotropic systems $\phi_c^x\equiv\phi_c^y$.

We simulate Boolean models with rectangles for different values of $\alpha$, that is, for different degrees of anisotropy.
Their aspect ratios ${b}/{a}$ are fixed within one model, but we simulate several models with different ratios ${b}/{a}\in\{1,\frac{3}{4},\frac{1}{2},\frac{1}{4}\}$ and anisotropy parameters $\alpha\in\{0,1,3,15,\infty\}$.
The models thus vary from isotropic orientation distributions to full alignment.
For each system, we determine the connectivity $C$ as a function of the intensity for differently large simulation boxes with linear system sizes $L\in\{40a, 50a, 60a, 70a, 75a, 80a, 85a, 95a\}$.
As discussed above, the more anisotropic a Boolean model the larger the simulation boxes must be so that the connectivity is accurately described by an error function (so that the finite size scaling can be applied).
For each system, we fit error functions separately to the both the connectivity in $x$- and in $y$-direction.
The fit is applied to the five largest system sizes which are simulated for the according Boolean model.
We thus estimate the effective percolation thresholds $\phi_c^{\mathrm{eff},x}$ and $\phi_c^{\mathrm{eff},y}$.
Next we extrapolate the effective percolation thresholds to the limit of infinite system size using (\ref{eq:scaling_xy}) with fit parameters $\phi_c^{x}$ and $m^{x}$ or $\phi_c^{y}$ and $m^{y}$.

Figure~\ref{fig_perco_scaling} shows for the six exemplary Boolean models the effective percolation thresholds $\phi_c^{\mathrm{eff},x}$ and $\phi_c^{\mathrm{eff},y}$ as a function of the rescaled system size.
It also plots the extrapolation and the corresponding bands of one or two standard deviations.
Within statistical significance, the percolation threshold is isotropic.
Even for the most anisotropic systems, we find $\phi_c^x\equiv\phi_c^y$.
Similarly, Balberg et al.~\cite{Balberg:1983, BalbergBinenbaumWagner1984} found isotropic percolation thresholds in two- and three-dimensional anisotropic systems of sticks.


This isotropy of the percolation threshold, which means that the Boolean model either percolates simultaneously in all directions or not at all, can be expected for Boolean models for at least three reasons:
First, consider the extreme case of aligned rectangles.
A dilation can transform a Boolean model with aligned squares into a Boolean model with aligned rectangles of arbitrary aspect ratios.
However, the affine transformation neither changes the area fraction nor the topology of the model.
Therefore, the percolation threshold for aligned rectangles must be independent of the aspect ratio.
Moreover, because the model can either be dilated in $x$- or $y$-direction, the percolation threshold must be isotropic.
Balberg and Binenbaum~\cite{Balberg:1983} provide a similar argument for the anisotropic systems of sticks.
They also discuss a renormalization-group argument why in an anisotropic square lattice an isotropic percolation threshold is expected.

Another plausibility argument for the isotropic percolation threshold is that, as indicated in Section~\ref{sec:percolation_connectivity}, percolation in $x$-direction is almost surely not possible along a straight line but only along a winded path.
This requires that also grains with different $y$-coordinates are connected.

The most fundamental reason for the isotropy of the percolation threshold is that it is closely related to the uniqueness of the percolating cluster in continuum percolation.
Meester and Roy~\cite{MeesterRoy1994} have proven for Boolean models of spheres with varying size that there can be at most one unbounded component, but the results can be generalized to grains with general shape, see also \cite{MeesterRoy1996}.
Either with probability one or zero there is either exactly one or no percolating cluster, in other words, the percolating cluster is (almost surely) unique.
Similar proofs could be derived for other percolation models~\cite[\eg,][]{AizenmanKestenNewman1987, BurtonMeester1993}.

\begin{figure}[t]
  \hspace*{0.16\textwidth}%
  \includegraphics[width=0.84\textwidth]{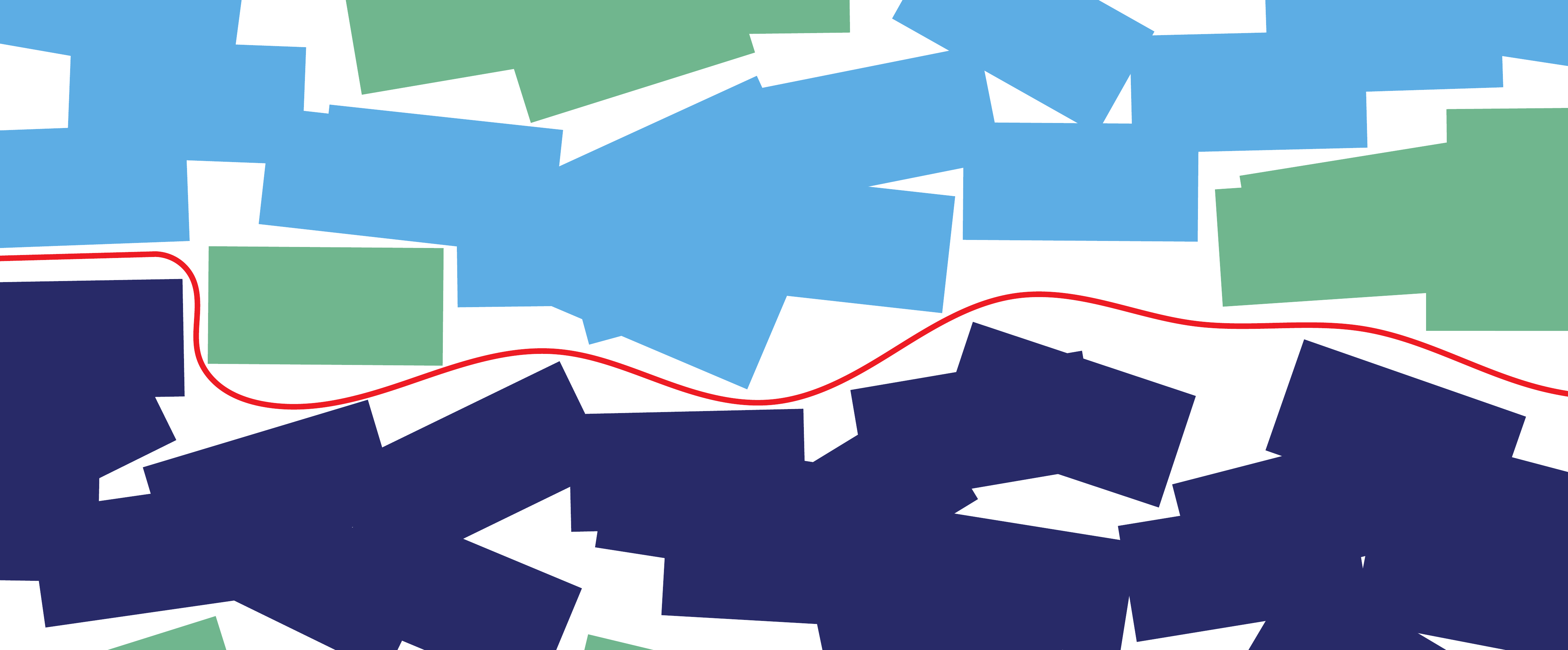}
  \caption{If an anisotropic model of overlapping grains percolated in $x$- but not in $y$-direction,
  there would appear with finite probability two percolating clusters (dark blue and cyan) not connected to each other (indicated by the red line that lies wholly in the void phase).}
  \label{fig:two_percolating_clusters}
\end{figure}

The isotropy of the percolation threshold follows by a heuristic argument from the uniqueness of the percolating cluster.
If we assume that the unbounded cluster spans the system only in $x$- but not in $y$-direction, two percolating clusters that are not connected to each other can appear with finite probability, see figure~\ref{fig:two_percolating_clusters}.
Assume that the percolating (dark blue) cluster is bounded in $y$-direction, that is, the cluster lies completely below some bound (maximum of red line).
Because of the statistical homogeneity of the system, the probability that there is another percolating cluster in the domain above the bound (colored cyan in figure~\ref{fig:two_percolating_clusters}) should be nonzero.
However, this would lead to a contradiction of the uniqueness of the unbounded cluster.
Therefore, we conjecture that in a statistically homogeneous system with almost surely at most one unique unbounded component this percolating cluster must be unbounded both in $x$- and in $y$-direction, and
the percolation thresholds must coincide:
\begin{eqnarray}
  \phi_c^x\equiv\phi_c^y.
\end{eqnarray}


\subsection{Threshold estimates for models with varying degrees of anisotropy}
\label{sec_perco_thresh_estimate}

The percolation threshold is by plausibility argument and empirical findings confirmed to be isotropic.
So the finite size scaling from (\ref{eq:scaling_xy}) can be replaced by
\begin{eqnarray}
    \phi_c^{\mathrm{eff},x} &= \phi_c + m^x\, L^{-1/\nu}\;\mathrm{and}\\
    \phi_c^{\mathrm{eff},y} &= \phi_c + m^y\, L^{-1/\nu}\;,
  \label{eq:scaling_xy_equiv}
\end{eqnarray}
where the percolation threshold in the infinite system $\phi_c$ is isotropic.
It is because of the nonuniversal prefactors $m^x$ and $m^y$ that the effective percolation thresholds differ in $x$- and $y$-direction.

\begin{figure}[t]
  \centering
  \includegraphics[width=\textwidth]{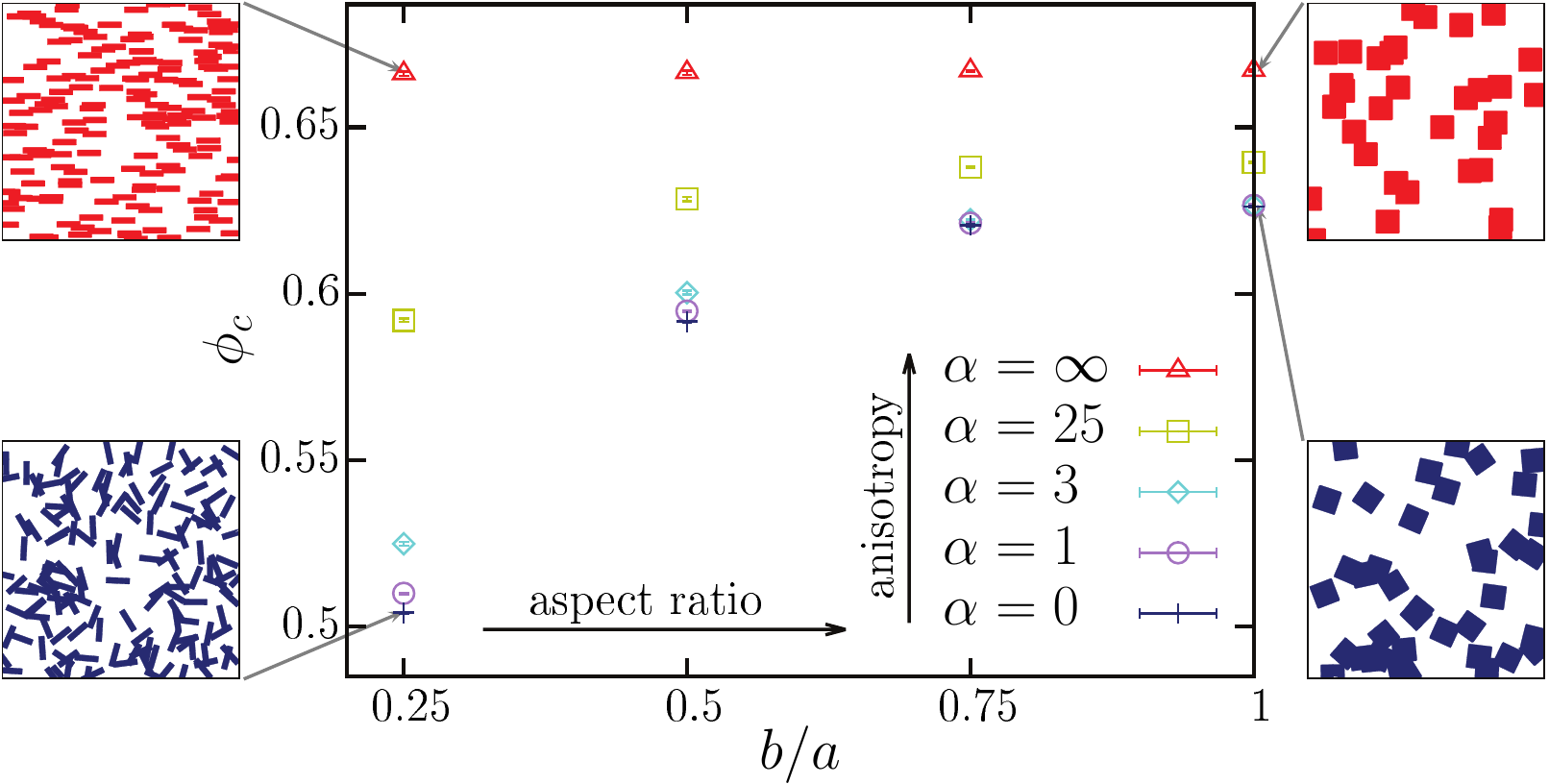}
  \caption{Isotropic percolation thresholds $\phi_c$ of Boolean models with orientation distributions with different degrees of anisotropy ($\alpha=0$ randomly oriented and $\alpha=\infty$ aligned rectangles) as a function of their aspect ratio ${b}/{a}$.
           Samples of rectangles (left) or squares (right) are depicted that are aligned (top) or randomly oriented (bottom).}
  \label{fig:percolation_thresholds_isotropic}
\end{figure}

We simultaneously extrapolate the effective percolation thresholds $\phi_c^{\mathrm{eff},x}$ and $\phi_c^{\mathrm{eff},y}$ to the same $\phi_c$ using a maximum likelihood fit described in \ref{sec:appendix}.
The final results for the isotropic percolation thresholds $\phi_c$ are plotted in figure~\ref{fig:percolation_thresholds_isotropic}.

If the grains are aligned, the percolation threshold does not depend on the aspect ratio of the grains, because in this special case changes in the aspect ratio of the single grains correspond to dilations of the Boolean model, as mentioned above.
The latter do not change the topology and therefore not the percolation threshold.
For all other orientation distributions of the grains, the percolation threshold decreases with a decreasing ratio of smaller to larger side length.
Put differently, the more elongated a rectangle with rotational degrees of freedom the smaller the percolation threshold.
In the limit of stick percolation, that is, for vanishing aspect ratio $b/a=0$, the percolation threshold vanishes as well $\phi_c=0$.
Comparing different orientation distribution, we find in agreement with previous findings~\cite{BalbergBinenbaumWagner1984}, the more anisotropic the orientation distribution, the larger is the percolation threshold.


\section{Approximations and bounds on percolation thresholds}
\label{sec:approx}

Numerical estimates of the percolation threshold can get computationally very expensive.
However, applications might need the threshold for some specific model parameters for which no simulation results are yet available.
A possibility to predict the percolation threshold without the need for expensive simulations would be of great importance; that is, an explicit formula approximating the percolation threshold.
Such an explicit formula should be easy computable and as accurate as possible~\cite{NeherMeckeWagner2008}.
An example is the scaling relation for varying dimensions from \cite{TorquatoJiao2013Dodecahedron},
which is based on a lower bound~\cite{Torquato2012} and a conjecture that no convex shape provides higher thresholds than the hyperspheres.

A common approximation is the excluded area approximation (or excluded volume in three dimensions)~\cite{Balberg:1984, BalbergBinenbaumWagner1984, Balberg:1985, Balberg:1987, NedaFlorianBrechet1999, WagnerBalbergKlein2006, AdlerEtAl2012, Chatterjee2015}.
It tries to predict percolation thresholds based on simulation results of similar models by approximating the dependence of the critical intensity on the model parameters by the change in the excluded area.
The excluded area for convex objects can be expressed explicitly by a so-called mixed intrinsic volume from integral geometry.
In section~\ref{sec:excluded-area-approx}, we compare the approximation of the excluded area to the percolation thresholds for the models with different degrees of anisotropy discussed above.

In contrast to the excluded area approximation, which uses an empirical parameter, Mecke and Wagner~\cite{MeckeWagner1991} suggested a purely geometrical approximation.
It uses an additive topological constant from integral geometry, the Euler characteristic.
For a compact set, it is given by the number of connected components minus the number of holes.
The critical intensity is approximated by the zero of its mean value (as a function of the intensity).
The Euler characteristic is positive for intensities below this zero and negative above.
Intuitively speaking, this indicates a change from a predominant role of single clusters to a network-like structure.
Explicit formulas for the mean Euler characteristic are known for very general Boolean models~\cite{Weil1990}, that potentially allow for estimates of the percolation threshold.
At least for the parametric Boolean models studied here, the approximation is a lower bound, see Section~\ref{sec:euler}.
We expect it to be a lower bound for even more general Boolean models, because the Euler characteristic does not distinguish holes in connected clusters from the percolating void phase.

Moreover, we find that the trend, \ie, the qualitative behavior, is very well described by the zero of the Euler characteristic.
For not too small aspect ratios of the rectangles, this finding allows us to very accurately estimate the percolation thresholds of Boolean models if the percolation threshold is known for squares that follow the same orientation distribution as the rectangles.

The Euler characteristic belongs to a class of so-called Minkowski functionals (or intrinsic volumes) from integral geometry~\cite{mecke_additivity_2000, SchneiderWeil2008, SchroederTurketal2010AdvMater}.
In two-dimensions, the two other Minkowski functionals are area and perimeter.
We investigate the potential of further approximations of the percolation threshold that do not use an empirical parameter, namely, the extremal points of the variances and covariances of the Minkowski functionals.
In Section~\ref{sec_percolation_cov}, we given an outlook on candidates for close upper bounds on the percolation thresholds.

\subsection{Excluded area approximation}
\label{sec:excluded-area-approx}

\begin{figure}[t]
  \hspace*{0.16\textwidth}%
  \includegraphics[width=0.4\textwidth]{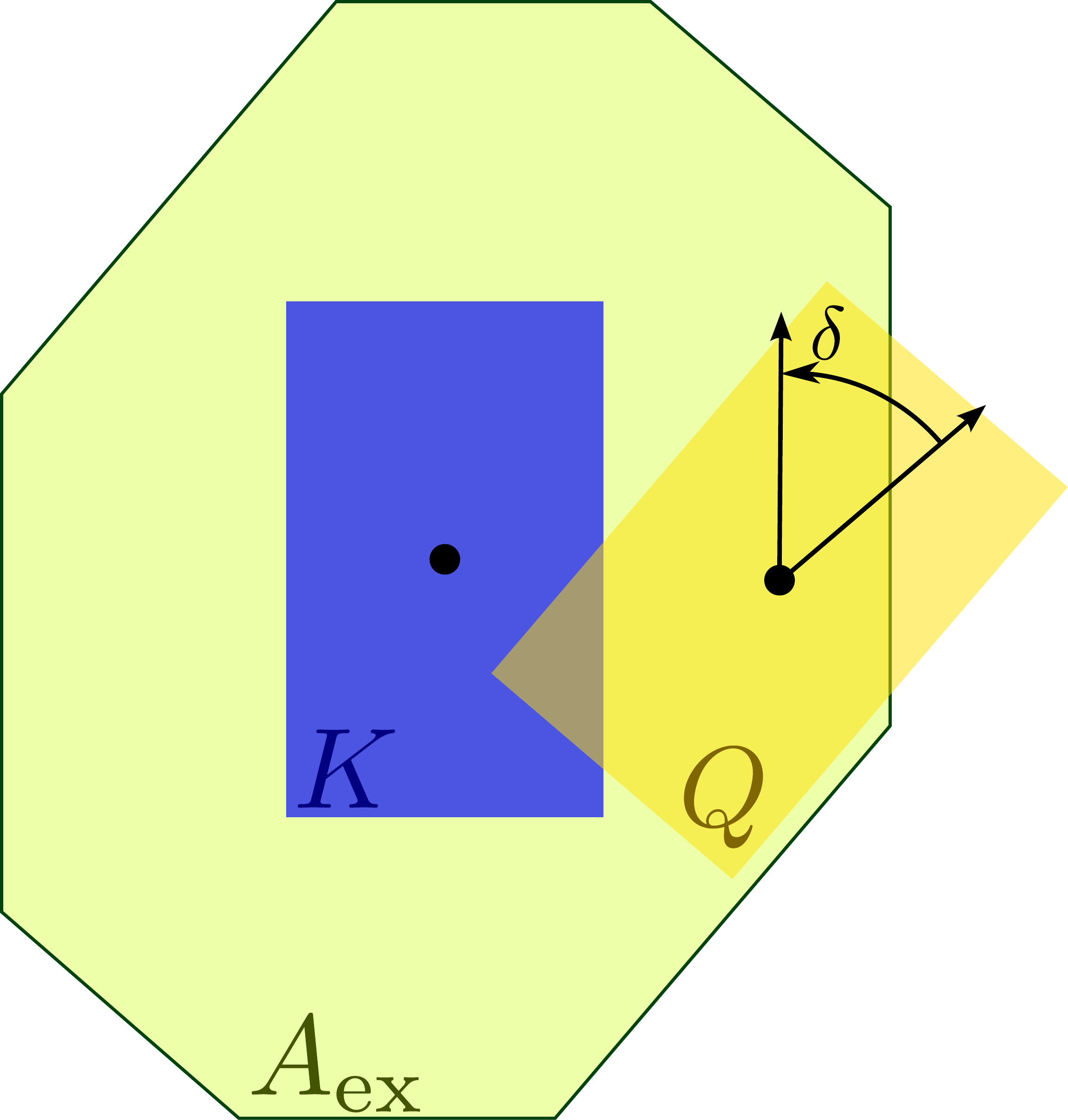}
  \caption{The excluded area $A_{\mathrm{ex}}(K,Q)$ of the two rectangles $K$ and $Q$ is the area of the green octagon: if the center of $Q$ is within this region, $K$ and $Q$ intersect.
$A_{\mathrm{ex}}(K,Q)=(a^2+b^2)\sin|\delta|+2ab\left(\cos|\delta|+1\right)$, where $\delta$ is the angle between the two rectangles, each with side lengths $a$ and $b$.}
  \label{fig_perco_A_ex}
\end{figure}

The excluded area is defined for two particles $K$ and $Q$.
It is the area of the region defined by all those center positions of the second particle $Q$ which leads to intersection of the two particles~\cite{Onsager1949}.
In the following, the number of grains that intersect an object is called its number of bonds.
In other words, the second particle $Q$ cannot enter this area without intersecting $K$.
Figure~\ref{fig_perco_A_ex} visualizes the excluded area of two rectangles.

The excluded area $A_{\mathrm{ex}}$ of two convex grains $K$ and $Q$ can be explicitly expressed using the so-called mixed functional $V^0_{1,1}$ from integral geometry~\cite{SchneiderWeil2008},
\begin{eqnarray}
  A_{\mathrm{ex}}(K,Q) = V^0_{1,1}(K,Q) + A(K) + A(Q)
  \label{eq:A_ex}
\end{eqnarray}
Note that our notation differs from \cite{Balberg:1984}, in which the same quantity is denoted by $\langle A \rangle$, but $\langle A_{\mathrm{ex}} \rangle$ indicates the total excluded area.
Averaging $A_{\mathrm{ex}}(K,Q)$ with the orientation distribution $\mathcal{P}(\theta)$ of the grains $K$ and $Q$ yields the averaged excluded area for single grains
\begin{eqnarray}
  \langle A_{\mathrm{ex}} \rangle =
  \iint_{-\frac{\pi}{2}}^{\frac{\pi}{2}}\mathrm{d}\theta_i\mathrm{d}\theta_j\,
  \mathcal{P}(\theta_i)\cdot \mathcal{P}(\theta_j)\cdot
  A_{\mathrm{ex}}(\vartheta(\theta_i)K,\vartheta(\theta_j)Q),
  \label{eq:avAex}
\end{eqnarray}
where $\vartheta(\theta)$ denotes a rotation by the angle $\theta$.

For the isotropic orientation distribution $\mathcal{P}(\theta)\equiv\mathrm{const.}$, the average mixed intrinsic volume separates.
In this special case, it is proportional to the product of the perimeters of the grains.
For arbitrary dimensions, a general formula for the exclusion volume of identical randomly oriented convex particles is given is given in \cite{TorquatoJiao2013Dodecahedron}. 
Therefore, the excluded area in two dimensions can, as it is well-known~\cite{Boublik1975}, be expressed by the areas $A$ and the perimeters $P$ of the single grains
\begin{eqnarray}
  \langle A_{\mathrm{ex}} \rangle = \frac{1}{2\pi}P(K)\cdot P(Q) + A(K) + A(Q).
  \label{eq:avAex_isotropic}
\end{eqnarray}

Also for rectangles with arbitrary orientations, there is an explicit formula for the mixed intrinsic volume in (\ref{eq:A_ex})~\cite[][p. 82]{HoerrmannEtAl2014}.
The excluded area of two rectangles $R$ with orientations $\theta_i$ and $\theta_j$ and side lengths $a$ and $b$ is given by
\begin{eqnarray}
  A_{\mathrm{ex}}(\vartheta(\theta_i)R,\vartheta(\theta_j)R)=(a^2+b^2)\sin|\theta_i-\theta_j|+2ab\left(\cos|\theta_i-\theta_j|+1\right).
  \label{eq_perco_Aex_rectangles}
\end{eqnarray}
This equation allows for an explicit expression approximating the percolation thresholds of Boolean models with rectangles.

The product of the critical intensity $\gamma_c$ and the average excluded area $\langle A_{\mathrm{ex}} \rangle$ is called the total excluded area.
It is equivalent to the average number of bonds per object at the critical intensity $B_c$, because the average number of bonds of a grain is equal to the average number of intersecting grains and thus equal to the average number of particles within the excluded area~\cite{Balberg:1984}
\begin{eqnarray}
  B_c = \gamma_c \cdot \langle A_{\mathrm{ex}} \rangle,
  \label{eq_perco_Bc_def}
\end{eqnarray}
which was confirmed numerically~\cite[\eg,][]{Balberg:1984, BalbergBinenbaumWagner1984, WagnerBalbergKlein2006}.

Balberg et al.~\cite{Balberg:1984} proposed that $B_c$ is, at least approximately, an invariant quantity for similar systems, following a similar argument from lattice percolation~\cite{ScherZallen1970}.
Actually, the average number of bonds per object $B_c$ is not a universal constant but depends on the grain shape and orientation distribution.
However, it varies to reasonable approximation only slightly for similar systems~\cite[\eg,][]{Balberg:1984, Baker:2002}.
If the average number of bonds per object at the critical intensity $B_c$ is empirically known (\eg, for a Boolean model of squares) and assumed to be the same for similar systems
(\eg, Boolean models with rectangles following the same orientation distribution) the critical intensity of these similar models can be approximated by the explicit expresssion
\begin{eqnarray}
  \gamma_c = \frac{B_c}{\langle A_{\mathrm{ex}} \rangle}.
  \label{eq_perco_Aex_approx}
\end{eqnarray}
In numerous simulation studies, the assumption of a system invariant total excluded area is shown to be a useful approximation~\cite{NedaFlorianBrechet1999, AdlerEtAl2012, Chatterjee2015} for which the critical intensity scales correctly with the particle size~\cite{WagnerBalbergKlein2006} and which approximates the functional dependence of the percolation threshold on the anisotropy of the model~\cite{Balberg:1984, BalbergBinenbaumWagner1984}.

The excluded area for overlapping rectangles with anisotropic orientation distribution is also calculated in~\cite{Chatterjee2015}, where the orientations are random within the interval $[-\alpha,\alpha]$ (following \cite{Balberg:1984}).
The approximation is compared to Monte Carlo simulations of isotropic models.
For increasing anisotropy, an increase of the percolation threshold is predicted, which we have confirmed by our simulations above.

Here we compare the excluded area approximation to the Monte Carlo estimates of the anisotropic parametric Boolean model studying the variety of models with different degrees of anisotropy from Section~\ref{sec:empirics}.

At the critical intensity, the average number of bonds $B_c$ for Boolean models can be derived from the percolation threshold $\phi_c$ using (\ref{eq:phi-gamma}) and (\ref{eq_perco_Bc_def}):
\begin{eqnarray}
  B_c = \frac{\langle A_{\mathrm{ex}} \rangle}{A}
  \ln\frac{1}{1-\phi_c}\;,
  \label{eq_perco_Bc_of_phic}
\end{eqnarray}
where $A$ is the area of a single grain.
Using the numerical results for the percolation thresholds $\phi_c$ from Section~\ref{sec:empirics}, we estimate $B_c$ for overlapping squares for the different orientation distributions.
Using (\ref{eq_perco_Aex_approx}), we finally approximate the critical intensities of the overlapping rectangles with varying aspect-ratio.

\begin{figure}[t]
  \centering
  \includegraphics[width=\textwidth]{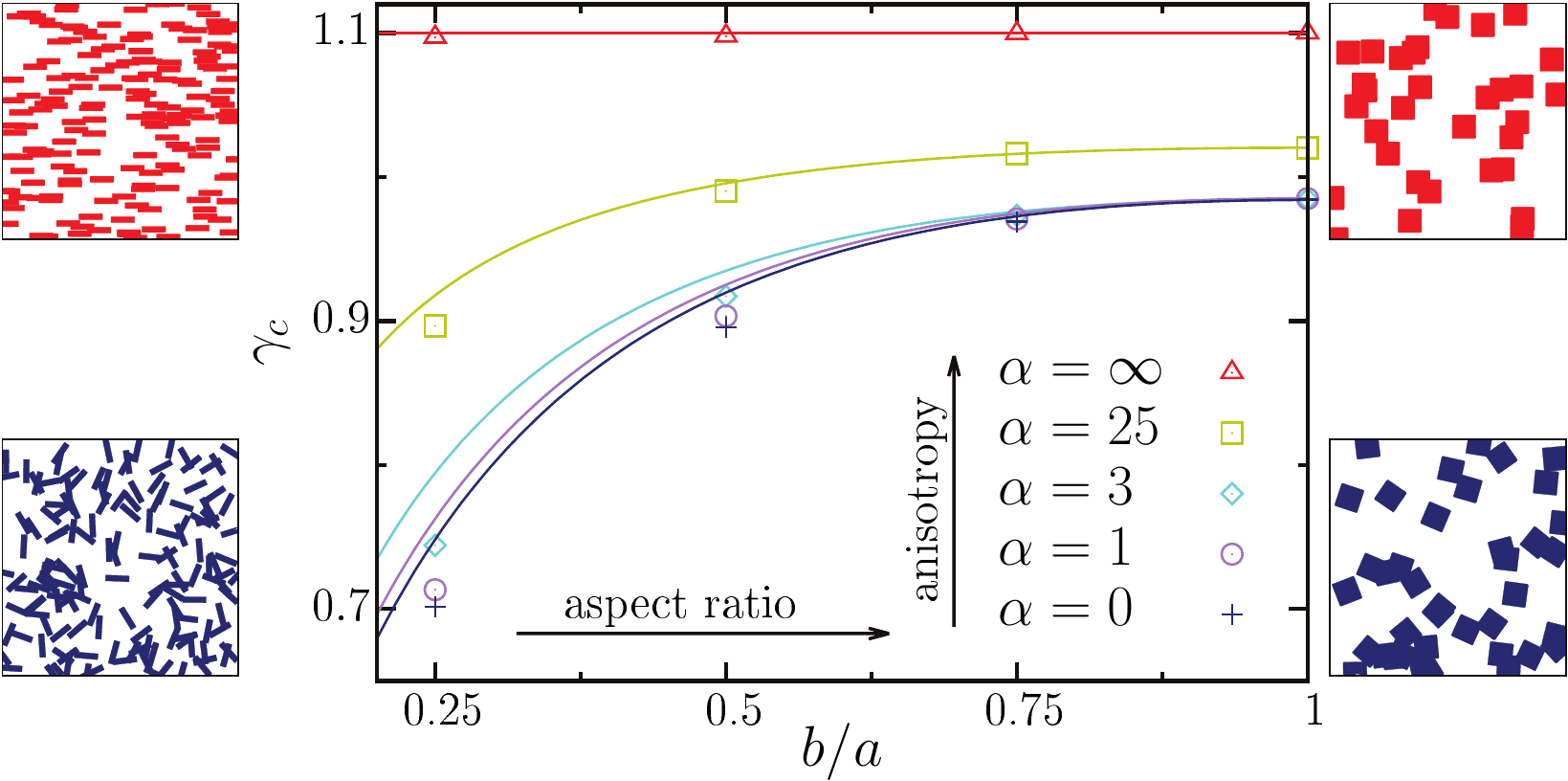}
  \caption{The excluded area approximation for the critical intensities $\gamma_c$ as a function of the aspect ratio ${b}/{a}$ of the rectangles for Boolean models with different degrees of anisotropy, see (\ref{eq_perco_Aex_approx}).
    A constant average number of bonds per object $B_c$ is assumed for the same orientation distribution, and the average excluded areas are calculated by (\ref{eq:avAex}) and (\ref{eq_perco_Aex_rectangles}).
    The estimates of $B_c$ are for each curve derived from the percolation thresholds of squares with the corresponding anisotropy parameter $\alpha$, see figure~\ref{fig:percolation_thresholds_isotropic} and (\ref{eq_perco_Bc_of_phic}).
    In other words, the critical intensities $\gamma_c$ of the squares are used as empirical parameters, so that the functional dependence of $\gamma_c$ on the aspect ratio can be approximated.
    At the left- and right-hand side samples of the Boolean models are depicted.}
  \label{fig:excluded_area_approximation}
\end{figure}

Figure~\ref{fig:excluded_area_approximation} compares this excluded area approximation to the simulation results from figure~\ref{fig:percolation_thresholds_isotropic}.
For aligned rectangles, the independence of the percolation thresholds from the aspect ratio is correctly reproduced, because the excluded area of aligned rectangles $A_{\mathrm{ex}}(R,R) = 4ab$ is proportional to the area of a single grain (which is here chosen to be unity).
For the other anisotropic or isotropic Boolean models, the excluded area approximation seems to be a close upper bound for the rectangles if the empirical parameters of a square following the same orientation distribution are used.
However, there are significant deviations from the prediction, for example, of the percolation threshold of randomly orientated rectangles with aspect ratio $\frac{1}{4}$.

\subsection{Euler characteristic approximation}
\label{sec:euler}

In contrast to the excluded area approximation that requires the input of an empirical parameter ($B_c$),
Mecke and Wagner~\cite{MeckeWagner1991} suggested a purely geometrical approximation, namely, the zero of the mean Euler characteristic as a function of the intensity.
They were motivated by the idea that a vanishing Euler characteristic implies a balance between clusters and holes.
Moreover, for self-dual matching lattices, like the square-lattice, the zero of the Euler characteristic and the critical point coincide exactly~\cite{SykesEssam1964}.
They compared the critical intensity $\gamma_c$ to the zero of the Euler characteristic $\gamma_0$, for example, for overlapping discs and spheres, and suggested that $\gamma_0$ could be a close bound for $\gamma_c$; see also~\cite{okun_euler_1990}.

Mecke and Seyfried~\cite{MeckeSeyfried2002} found in Monte Carlo simulations of a large class of discretized Boolean models a good qualitative agreement, in that the dependence on the shape of the constituents is captured well, but they found also quantitative deviations.
Neher et al.~\cite{NeherMeckeWagner2008} analyzed two-dimensional lattice graphs, for which the Euler characteristic provides a good approximation of the critical point.
The criterion has, for example, been applied to study the percolation in level sets of Gaussian random fields~\cite{tomita_percolation_1994, roubin_critical_2016}, biopolymeric networks~\cite{mickel_robust_2008}, flow in heterogeneous soil~\cite{neuweiler_upscaling_2007}, or surface roughness of medical implants~\cite{RodriguezEtAl2010}.
Moreover, the connection between percolation and the Euler characteristic triggered predictions of flow in porous media that are based on the Minkowski functionals~\cite{ScholzEtAl2012, scholz_direct_2015}.

More precisely, we here refer to the density of the Euler characteristic $\overline{\chi}$, that is, before taking the limit of an infinitely large observation window the functional value is rescaled by the size of the window~\cite{SchneiderWeil2008}.
Explicit formulas are known for the density of the Euler characteristic of Boolean models as a function of the intensity~\cite{Weil1990}. 
The criterion is thus easily applicable to a great variety of continuum percolation models.

\begin{figure}[t]
  \centering
  \includegraphics[width=\textwidth]{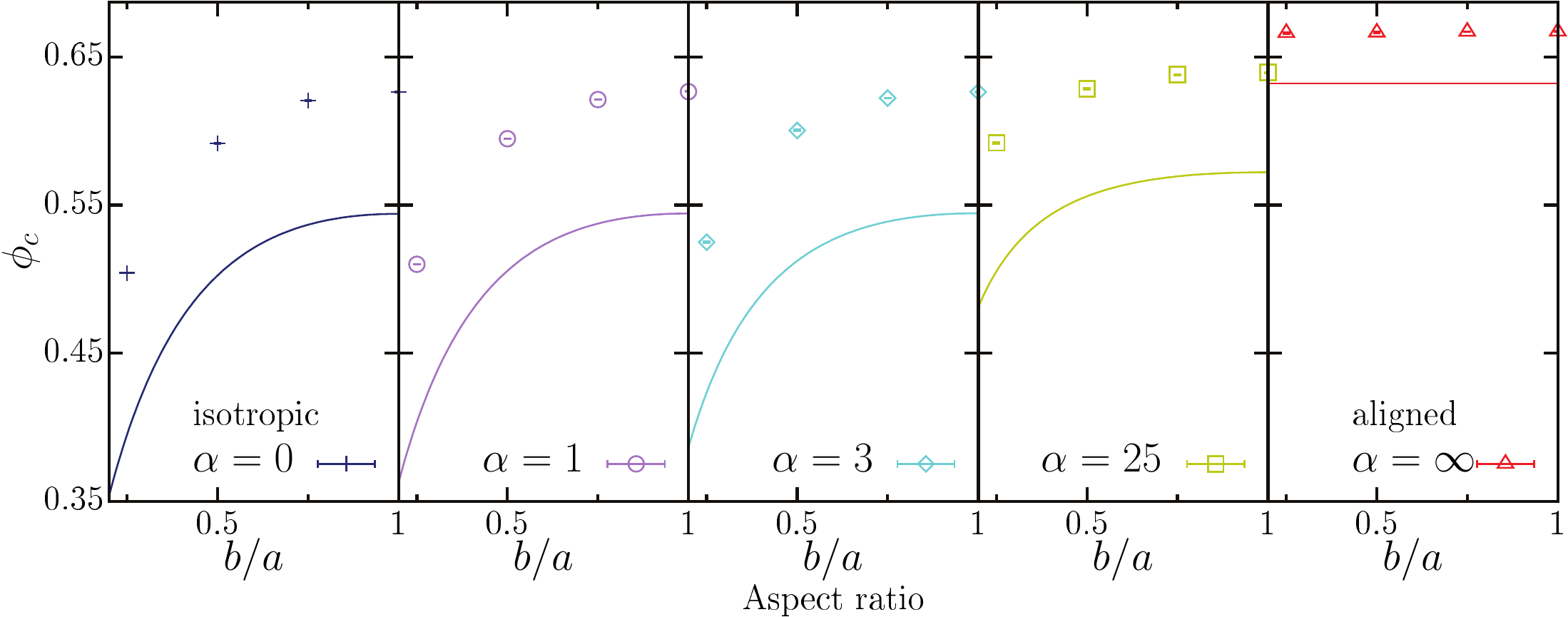}
  \caption{The percolation thresholds $\phi_c$ (marks) are approximated by the zero of the Euler characteristic (lines) for Boolean models with rectangles.
The single plots show the percolation thresholds as a function of the aspect ratio for orientation distributions with different degrees of anisotropy, from randomly oriented (far left) to aligned (far right) rectangles.
The zero of the Euler characteristic is a lower bound on the percolation thresholds.
 Although there is a significant offset, the qualitative behavior, \ie, the dependence on anisotropy and grain shape, is described quite accurately.}
  \label{fig_perco_Euler_zero}
\end{figure}

Here we calculate this approximation for the Boolean models with varying anisotropy that are studied above,
and we compare it to the simulation results for the percolation thresholds from Section~\ref{sec:empirics}.
We investigate how well the dependence of the threshold on the grain shape and orientation distribution is captured by the zero of the Euler characteristic.
The density $\overline{\chi}$ of the Euler characteristic of a Boolean model with rectangles as a function of the intensity $\gamma$ is given by
\begin{equation}
  \overline{\chi}(\gamma) = \gamma \left(1 - \frac{\gamma}{2}\langle{V}^0_{1,1}\rangle\right)\exp(-\gamma\,A),
\end{equation}
see \cite{Weil1990}, where $\langle{V}^0_{1,1}\rangle$ is the 
average of the mixed intrinsic volume ${V}^0_{1,1}$ of two rectangles,
\begin{eqnarray}
  \langle{V}^0_{1,1}\rangle :=
  \iint_{-\frac{\pi}{2}}^{\frac{\pi}{2}}\mathrm{d}\theta_i\mathrm{d}\theta_j\,
  \mathcal{P}(\theta_i)\cdot \mathcal{P}(\theta_j)\cdot
  V^0_{1,1}(\vartheta(\theta_i)R,\vartheta(\theta_j)R),
\end{eqnarray}
see (\ref{eq:A_ex}) and (\ref{eq:avAex}).
The zero of the Euler characteristic then follows in a straightforward way:
\begin{eqnarray}
                     & \chi(\gamma_0)                                  &= 0 \\
    \Leftrightarrow \quad & 1 - \frac{\gamma_0}{2}\langle{V}^0_{1,1}\rangle &= 0 \\
    \Leftrightarrow \quad & \gamma_0                                        &= \frac{2}{\langle{V}^0_{1,1}\rangle},
  \label{eq_perco_zero_EC}
\end{eqnarray}
which appears similar to the excluded area approximation in (\ref{eq_perco_Aex_approx}).
The occupied area fraction $\phi_0$ at which the mean Euler characteristic changes its sign is according to (\ref{eq:phi-gamma}) then given by
\begin{eqnarray}
  \phi_0 := 1-\mathrm{e}^{-\gamma_0 A}\;,
  \label{eq_perco_phi0}
\end{eqnarray}
where $A$ is the area of a single grain.

\begin{figure}[t]
  \centering
  \includegraphics[width=\textwidth]{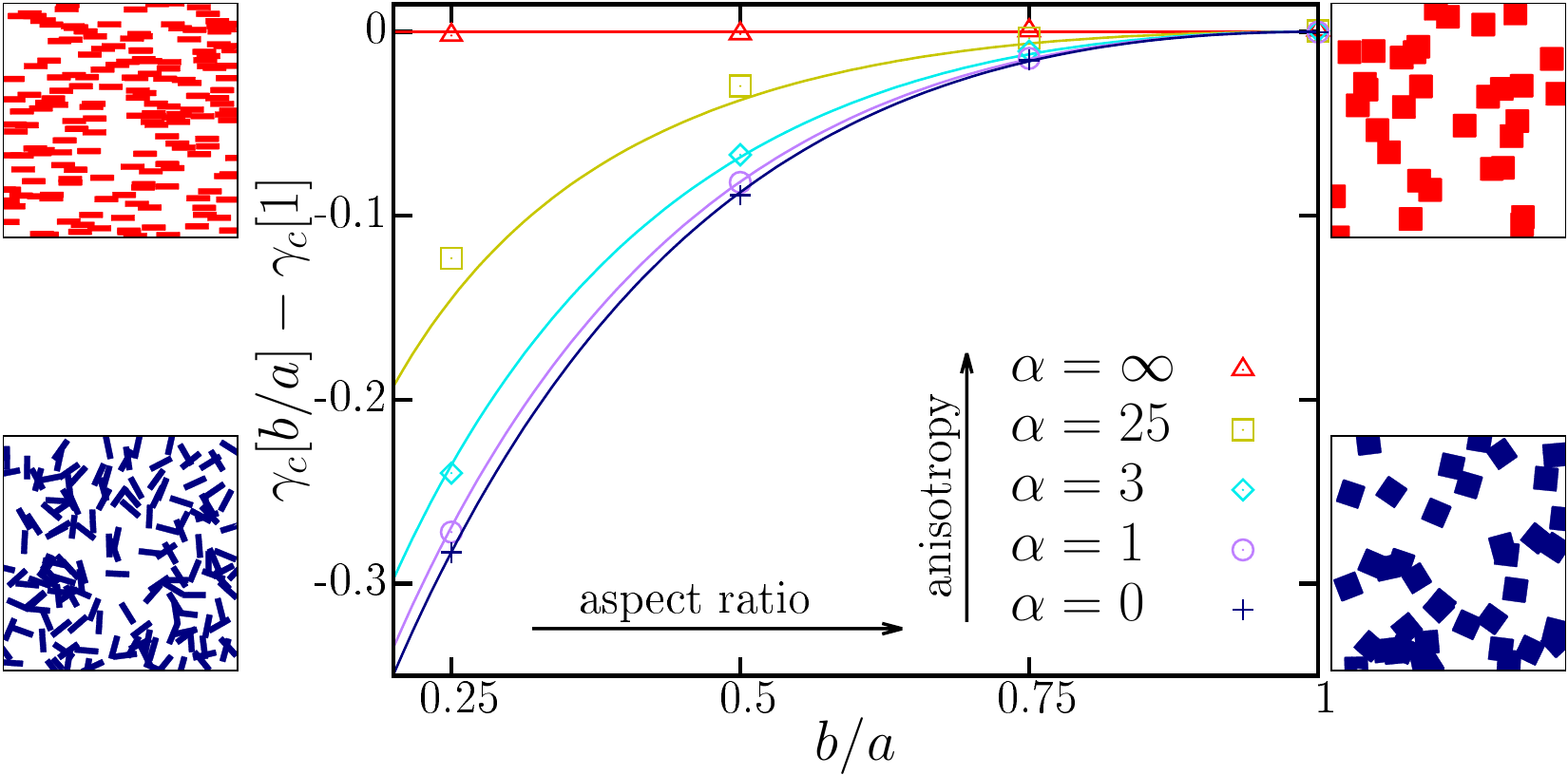}
  \caption{The change in the critical intensity $\gamma_c$ (marks) for different aspect ratios ${b}/{a}$ is well described by the change in the zero of the Euler characteristic $\gamma_0$ (lines); not only in the trivial case of aligned rectangles (red, see also samples depicted at the top of the right- and left-hand side), but a remarkably good approximation is also provided for randomly oriented rectangles with aspect ratios ${b}/{a}>\frac{1}{4}$ (blue, see also samples depicted at the bottom of the right- and left-hand side).
For details, see figure~\ref{fig:percolation_thresholds_isotropic}.}
  \label{fig:offset_EulerZero}
\end{figure}

Figure~\ref{fig_perco_Euler_zero} compares $\phi_0$ from (\ref{eq_perco_phi0}) to the estimates of the isotropic percolation thresholds $\phi_c$ from figure~\ref{fig:percolation_thresholds_isotropic}.
At least for these Boolean models, the zero of the Euler characteristic is a lower bound on $\phi_c$.
There is a significant quantitative deviation, but similar to the findings for discretized Boolean models~\cite{MeckeSeyfried2002}, we find that the qualitative behavior of $\phi_c$, \ie, the dependence on anisotropy and grain shape, is well described by $\phi_0$.

To further investigate this accurate qualitative description, we plot in figure~\ref{fig:offset_EulerZero} the difference of the critical intensity of rectangles and the critical intensity of squares with the same orientation distribution.
We compare these differences to the differences of the corresponding zeros of the Euler characteristic.
For the Boolean models studied here, we find a remarkable agreement.
Only for an intermediate regime of strongly preferred orientation, but no full alignment ($\alpha=25$), we find distinct (but still small) deviations.
However, there is an excellent agreement of the difference in the critical intensities to the difference of the zeros of the Euler characteristic not only for the special case of full alignment but also for randomly oriented rectangles.
If the critical intensity of squares is used as an empirical parameter, similar to the excluded area approximation, the zero of the Euler characteristic allows at least for these Boolean models a quite precise prediction of the percolation threshold, compare figure~\ref{fig:excluded_area_approximation}.

Like the critical average number of bonds $B_c$, the offset between the critical intensity $\gamma_c$ and the zero $\gamma_0$ of the Euler characteristic depends on the shape of the grains; it is different for randomly oriented squares, 0.19688(1)~\cite{JiantongOestling2013}, and discs, 0.128085(2)~\cite{QuintanillaZiff2007}.
Note, however, that assuming a constant offset to the number of grains per area of a single grain would lead to a nonphysical behavior in the limit of vanishing aspect ratios.
For grains with unit area, both the critical intensity and the zero of the Euler characteristic trivially vanish in the limit of extreme anisotropy, \ie, for sticks, because of the choice of units.
For example, for randomly oriented sticks of length unity, the zero of the Euler characteristic ($\gamma_0=\pi$) is as expected a lower bound on the critical intensity ($\gamma_c=5.6372858(6)$~\cite{LiZhang2009}).

\subsection{Second-moment approximations}
\label{sec_percolation_cov}

The Euler characteristic is one of the so-called Minkowski functionals (or intrinsic volumes) from integral geometry~\cite{mecke_additivity_2000, SchneiderWeil2008, SchroederTurketal2010AdvMater}.
In two-dimensions, the Minkowski functionals are the area $W_0$, the perimeter $W_1$, and a functional proportional to the Euler characteristic $W_2=2\pi\chi$.
These three Minkowski functionals characterize, according to Hadwiger's theorem, the complete ``additive shape information'' of a compact and convex set $K$ in the following sense:
they form a basis of all functionals $F$ that are motion invariant, continuous, and additive~\cite{Hadwiger1957, SchneiderWeil2008}.
The latter means that $F(K\cup L) = F(K)+F(L)$ for two disjoint sets $K$ and $L$.
The second moments of Minkowski functionals and generalizations thereof are known analytically~\cite{kerscher_morphological_2001, brodatzki_simulating_2002, HugLastSchulte2013, hug_second_2016}.
Remarkably, a behavior similar to thermodynamical quantities in statistical physics has been found for the second moments of Minkowski functionals~\cite{mecke_additivity_2000, mecke_exact_2001}.

As an outlook, we here examine the prospects of further approximations of the percolation threshold based on the second moments of the Minkowski functionals without using empirical parameters.
We compare the extremal points of the variances and covariances of the Minkowski functionals to the empirical estimates of the critical intensity.

An interesting example, which can serve as a motivation of this approach, is cell percolation in planar Poisson Voronoi tessellations.
Voronoi cells of a Poisson point process are colored black with some probability $p$, and a percolating cluster is an unbounded and connected collection of black cells.
The critical probability of this model of continuum percolation is $p_c=\frac{1}{2}$~\cite{BollobasRiordan2006}.
Last and Ochsenreither~\cite{LastOchsenreither2014} have proven that this coincides with the global maximum of the asymptotic variance of the Euler characteristic.
Moreover, Hug et al.~\cite{HugLastSchulte2013} found for overlapping discs that both the local minimum of the variance of the perimeter and the local minimum of the covariance of area and Euler characteristic are good approximations of the percolation threshold.
For example, the local minimum of the asymptotic covariance $\sigma_{0,2}$ is found at $\gamma\approx 1.15$ compared to the critical intensity $\gamma_c\approx 1.13$~\cite{Quintanilla:2000}, where the unit of area is again defined by the grain.

\begin{figure}[t]
  \centering
  \includegraphics[width=0.48\textwidth]{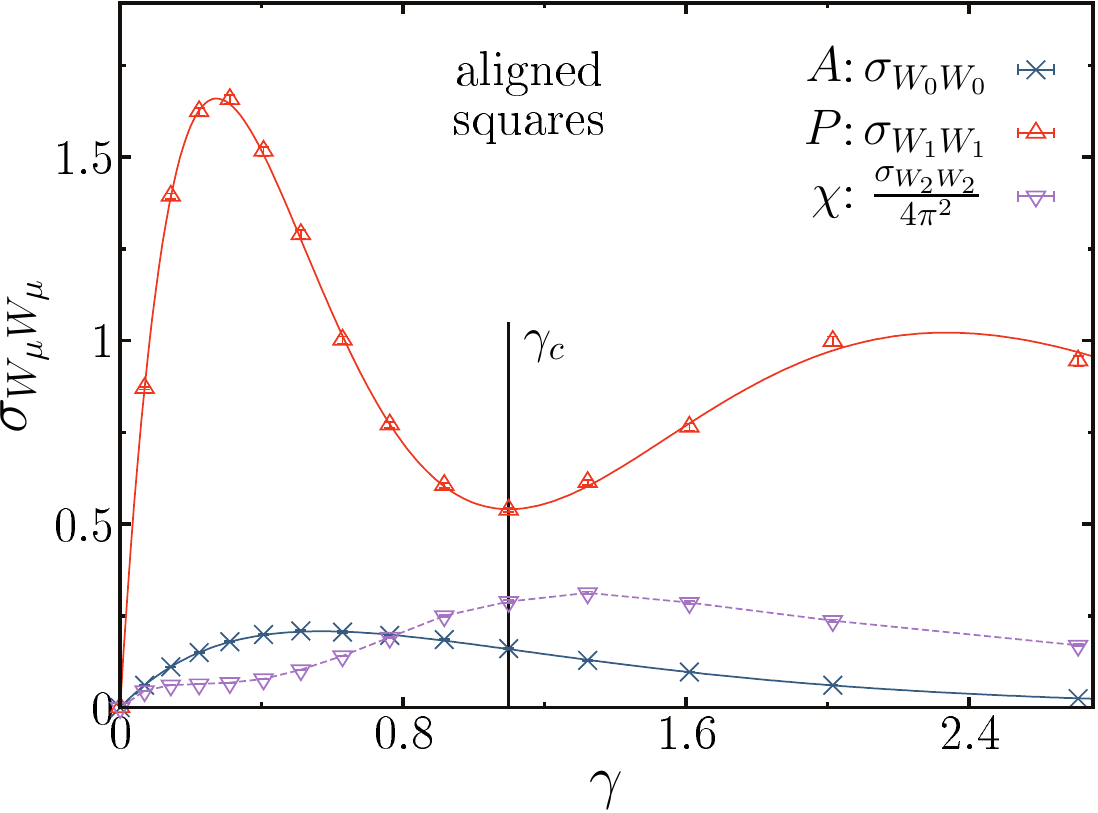}
  \hfill
  \includegraphics[width=0.48\textwidth]{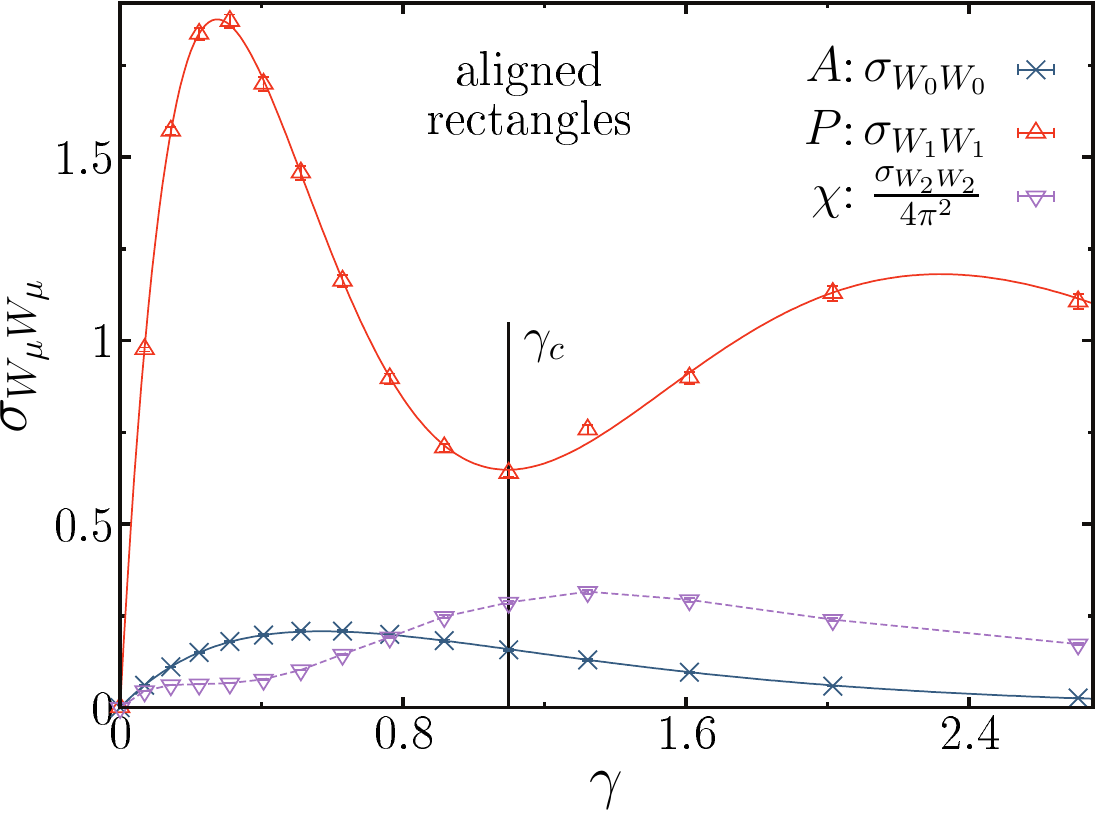}\\
  \includegraphics[width=0.48\textwidth]{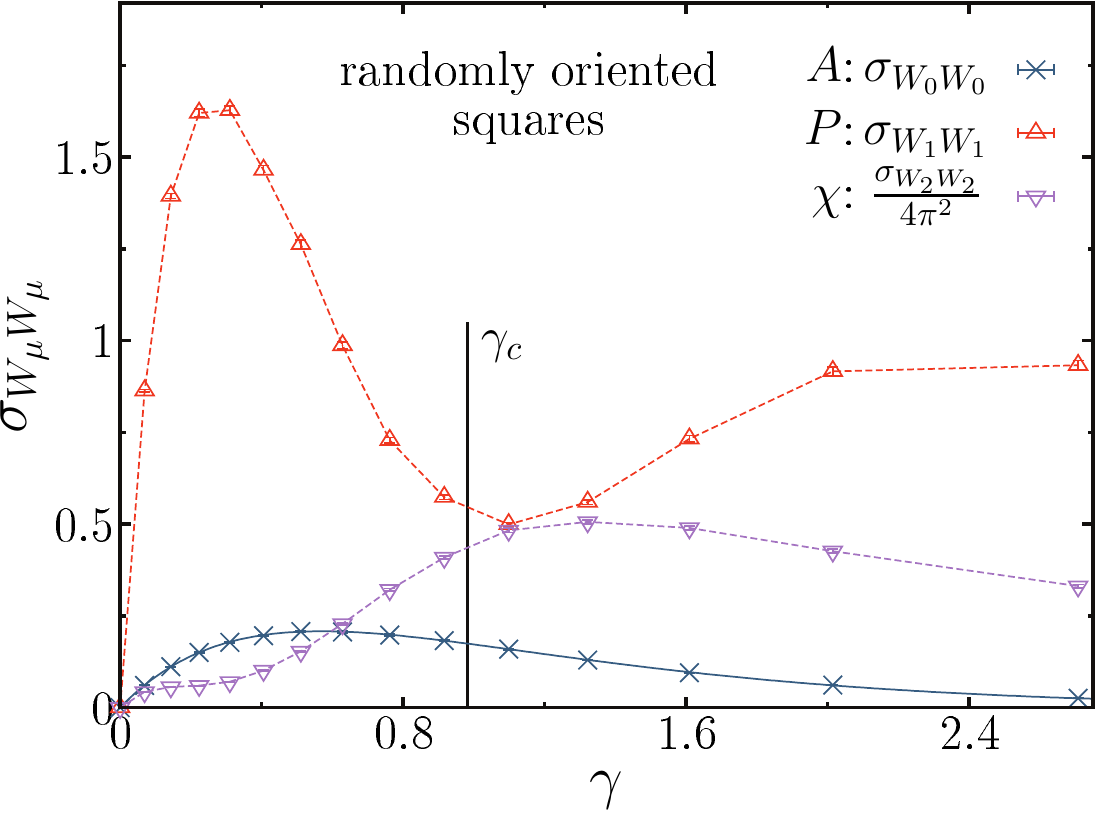}
  \hfill
  \includegraphics[width=0.48\textwidth]{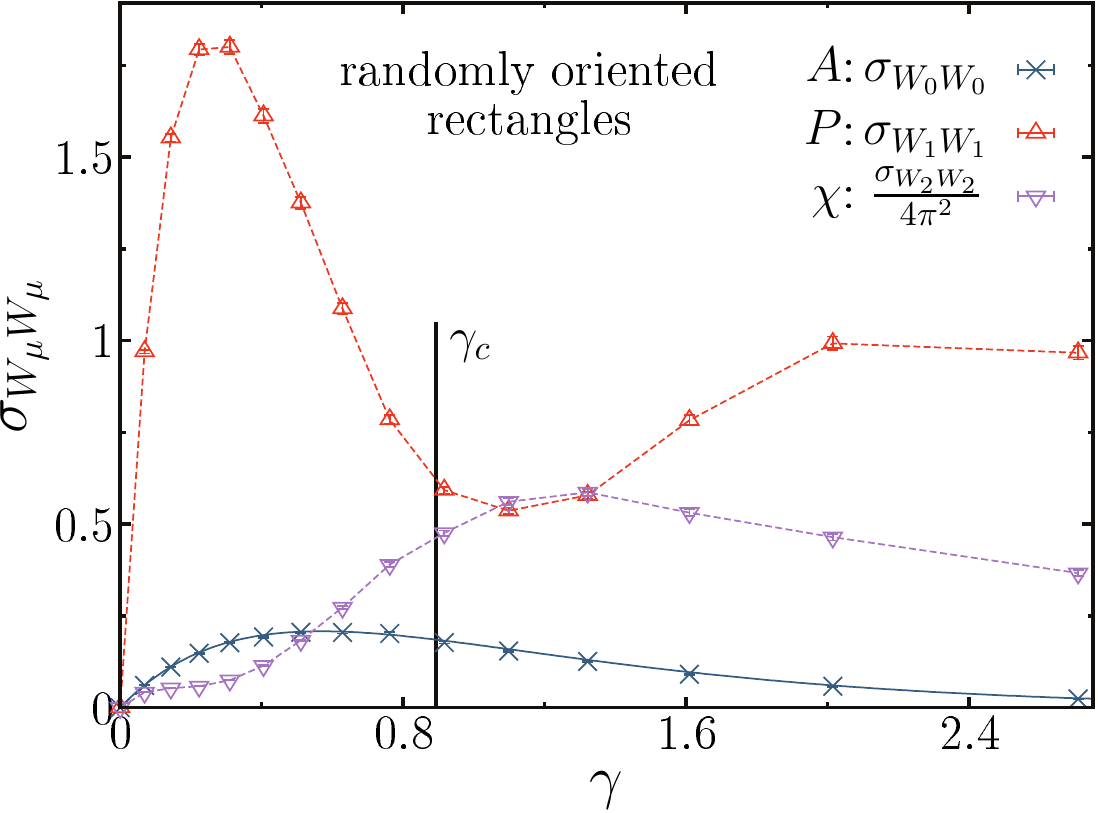}
  \caption{Asymptotic variances $\sigma_{W_{\mu}W_{\mu}}$ of the Minkowski functionals as functions of the intensity $\gamma$ of the Boolean models with squares (left)
  or rectangles with aspect ration $\frac{1}{2}$ (right), the rectangles are either fully aligned (top) or follow an isotropic orientation distribution (bottom).
  The marks represent numerical estimates from simulations, the solid lines represent the analytic curves, and the dashed lines are guides to the eye.
  The critical intensity $\gamma_c$ is indicated by the black vertical line, the estimate for the aligned squares is taken from \cite{TorquatoJiao2013Dodecahedron}
  and for the isotropic squares and rectangles from \cite{Baker:2002} or \cite{JiantongOestling2013}, respectively.}
  \label{fig_perco_var}
\end{figure}

\begin{figure}[t]
  \centering
  \includegraphics[width=0.48\textwidth]{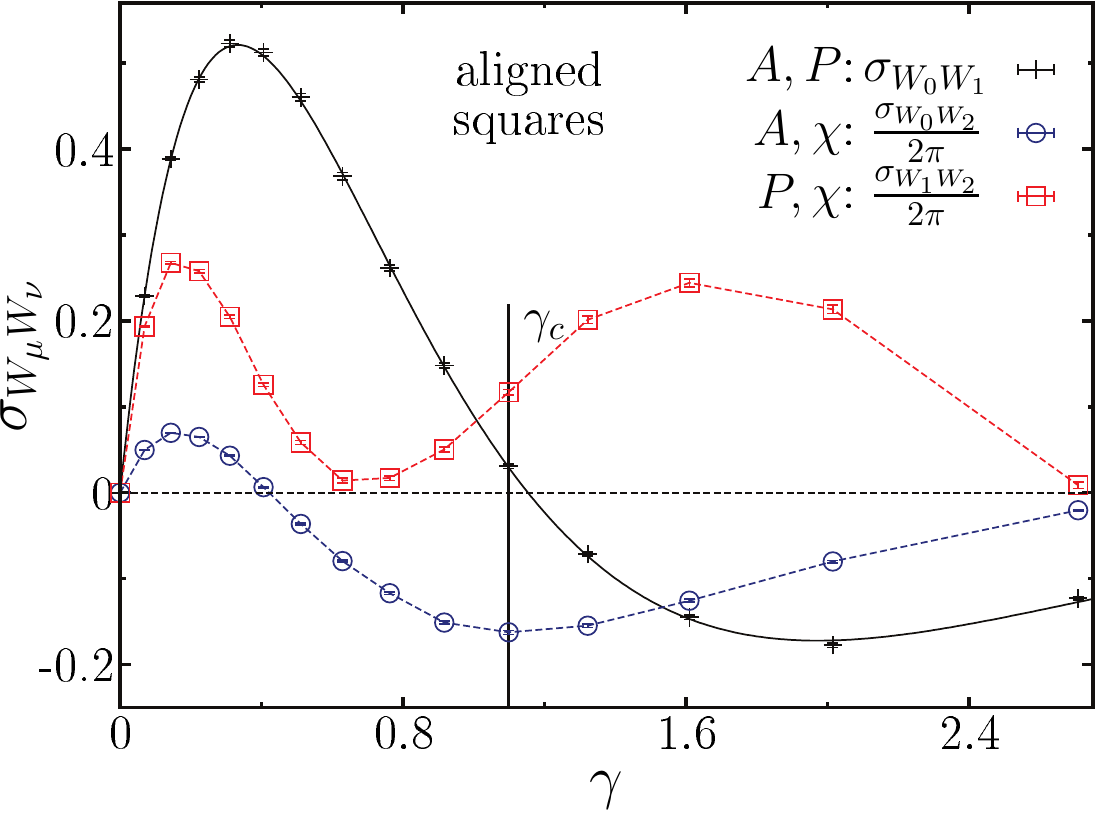}
  \hfill
  \includegraphics[width=0.48\textwidth]{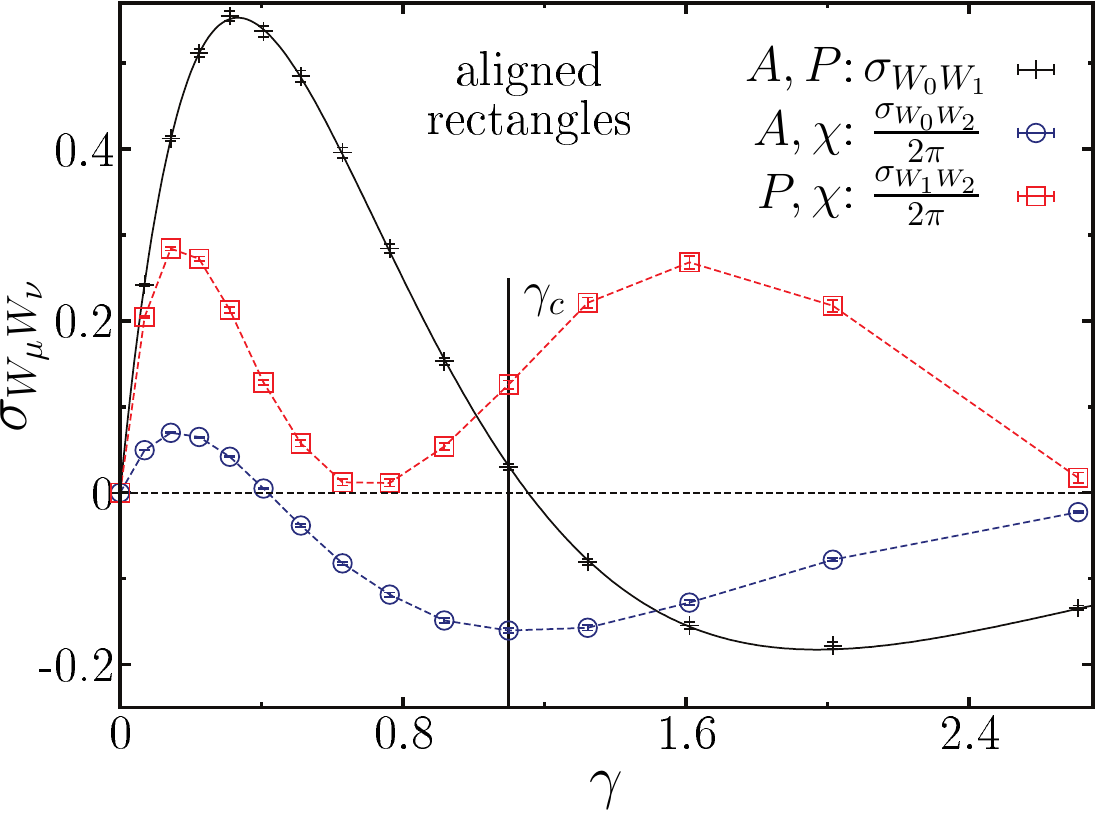}\\
  \includegraphics[width=0.48\textwidth]{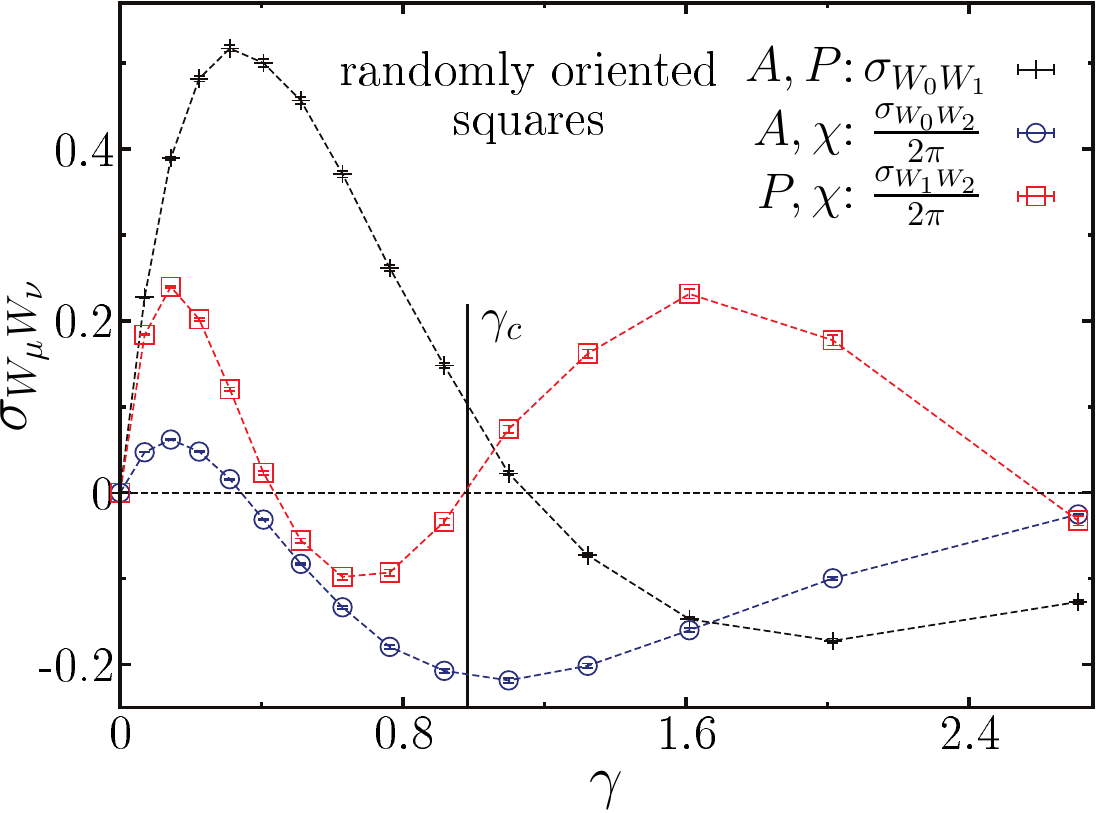}
  \hfill
  \includegraphics[width=0.48\textwidth]{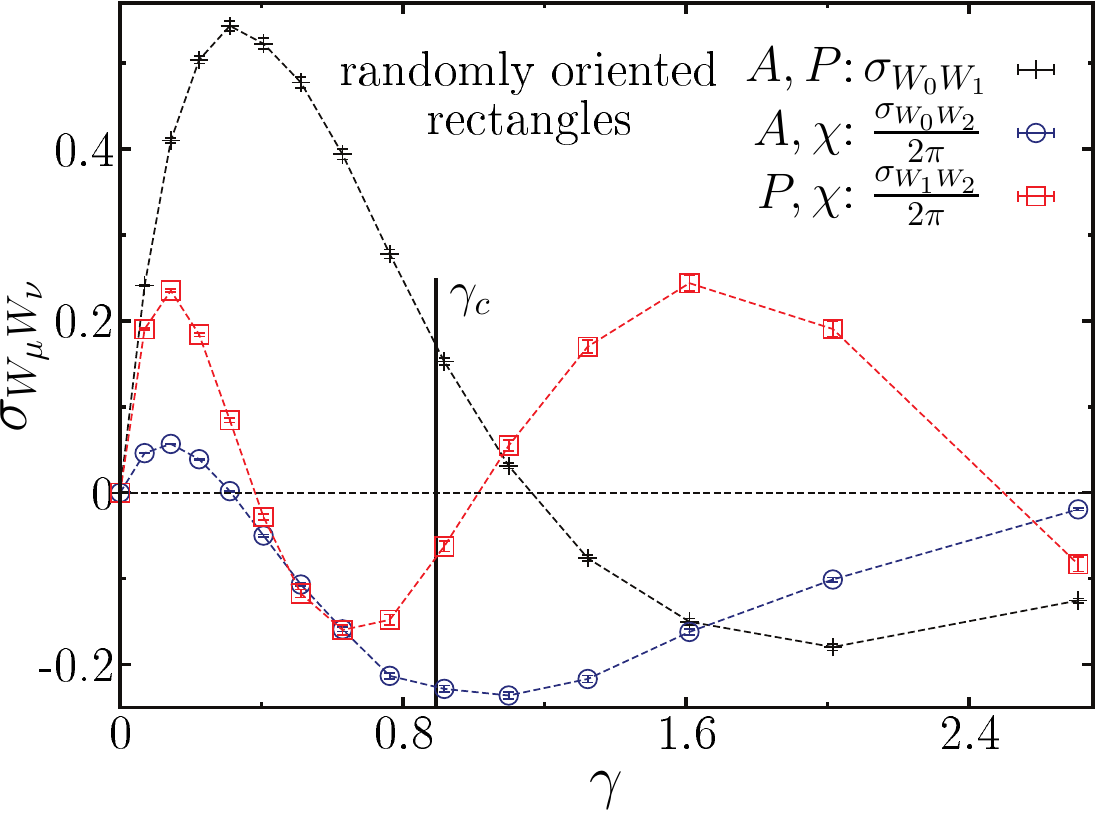}
  \caption{Asymptotic covariances $\sigma_{W_{\mu}W_{\nu}}$ of the Minkowski functionals as functions of the intensity $\gamma$ of the Boolean models with squares (left)
  or rectangles with aspect ration $\frac{1}{2}$ (right), the rectangles are either fully aligned (top) or follow an isotropic orientation distribution (bottom).
  For details, see figure~\ref{fig_perco_var}.}
  \label{fig_perco_cov}
\end{figure}

In our outlook here, we compare for four different Boolean models with rectangles the extremal points of the variance-covariance structure of the Minkowski functionals to the percolation threshold.
We investigate whether the second moments could allow for accurate predictions of the thresholds without the need for empirical parameters, especially testing whether the local minima of $\sigma_{0,2}$ or $\sigma_{1,1}$ are close to the percolation threshold for anisotropic models.

Figures~\ref{fig_perco_var} and \ref{fig_perco_cov} show the different asymptotic variances or covariances for a Boolean models with either squares or rectangles (with aspect ratio $\frac{1}{2}$).
These are either fully aligned or follow an isotropic orientation distribution.
For these four Boolean models very accurate estimates of the percolation thresholds can be found in literature~\cite{Baker:2002, mertens_continuum_2012, TorquatoJiao2013Dodecahedron, JiantongOestling2013}.
We use these precise estimates with at least four significant number of digits because we indeed find some remarkable agreement between the extremal points and the percolation threshold.

Like for the overlapping discs, a local minimum point of the variance of the perimeter is an excellent approximation of the percolation threshold for aligned squares or rectangles with aspect ratio $\frac{1}{2}$.
The critical intensity of aligned squares is $\gamma_c = 1.0982(3)$~\cite{TorquatoJiao2013Dodecahedron}, the local minimum point of the asymptotic variance of the perimeter is $\gamma_m[{b}/{a}=1] \approx 1.0994$, which is derived by numerically minimizing $\sigma_{W_1W_1}$ using \cite{SciPy}.
While this is a remarkably good approximation, there is a statistically significant difference.
Moreover, while the critical intensity must not depend on the aspect ratio of the aligned rectangles, see Section~\ref{sec_iso_perco}, the local minimum point of the asymptotic variance of the perimeter slightly depends on the aspect ratio.
For an aspect ratio of $\frac{1}{2}$, the local minimum point is at $\gamma_m[{b}/{a}=\frac{1}{2}] \approx 1.0952$, and for an aspect ratio of $\frac{1}{10}$, it is $\gamma_m[{b}/{a}=\frac{1}{10}] \approx 1.0727$.
The local minimum point of the variance is also neither an upper nor a lower bound for the percolation threshold.
For aligned rectangles, the minimum points can be below the critical intensity.
However, the minimum points are above the critical intensities for aligned squares or for the randomly oriented squares and rectangles, see figure~\ref{fig_perco_var}.

Another candidate for an approximation of the percolation threshold is the local minimum point of the covariance of area and Euler characteristic.
Figure~\ref{fig_perco_cov} shows that it is a very good approximation for aligned squares and rectangles with aspect ratio $\frac{1}{2}$ and a close upper bound if the grains are randomly oriented.

Only for one of the Boolean models, a vanishing correlation can serve as a good approximation of the percolation threshold: for randomly oriented squares the first zero of the covariance of perimeter and Euler characteristic ($\propto\sigma_{W_{1}W_{2}}$) seems to be in good agreement with the critical intensity, see figure~\ref{fig_perco_cov}.
However, for randomly oriented rectangles there is a distinct difference.
Moreover, for aligned squares or rectangles, as well as for overlapping discs, this zero crossing from negative to positive correlation does not exist.

\section{Conclusion and Outlook}
\label{sec_conclusion_and_outlook}

We have investigated percolation in anisotropic systems, both for finite simulation boxes and in the limit of infinite system size.
We have especially elucidated how the anisotropy of the Boolean model, which is a geometric property, influences percolation, which is a topological property.
There is both universal and non-universal behavior. 
The isotropy of the percolation threshold is of the first kind.
In infinite systems, there is no difference in effective percolation thresholds for different directions.
Independent of the details of the model, the difference between the effective percolation thresholds observed in finite systems vanishes in the thermodynamic limit.
This has here been related to the uniqueness of the percolating cluster, see figure~\ref{fig:two_percolating_clusters}.

In a finite sample of an anisotropic Boolean model, a cluster is more likely to span the system in the preferred than in the perpendicular direction, see figure~\ref{fig_perco_sample}.
This anisotropic percolation in finite samples is demonstrated by determining the effective percolation thresholds, see (\ref{eq_perco_def_phieff}) and (\ref{eq:erffit}), by a fit of an error function to the connectivity, see figure~\ref{fig:connectivity}.
Although a distinct difference remains even for very large system sizes, the extrapolation to infinite system sizes shows that even the most anisotropic model percolates simultaneously in all directions, see figure~\ref{fig_perco_scaling}.
This finding we here refer to as the isotropy of the percolation threshold.
Our results are in agreement with previous findings~\cite{Balberg:1983, BalbergBinenbaumWagner1984, BoudvilleMcGill1989, kale_effect_2016}.

For Boolean models with different degrees of anisotropy, we have estimated the isotropic percolation thresholds by a simultaneous fit of the finite size scaling in $x$- and $y$-direction, see (\ref{eq:scaling_xy_equiv})--(\ref{eq:percolation_fit}).
While the isotropy of the percolation threshold is universal, its value is non-universal, that is, it depends on the grain distribution (or the degree of orientation bias).
In agreement with previous findings~\cite{BalbergBinenbaumWagner1984}, the more anisotropic the orientation distribution, the larger is the percolation threshold, see figure~\ref{fig:percolation_thresholds_isotropic}.

There are efficient algorithms that use inhomogeneous Boolean models for high-precision estimates of percolation thresholds in homogeneous and isotropic models, \eg, see \cite{Quintanilla:1999, Quintanilla:2000, LorenzZiff2001}.
The isotropy of the percolation threshold implies that these algorithms can be directly generalized to anisotropic systems.
Different directions of the gradient in the intensity should yield (within statistical errors) the same estimate of the percolation threshold. 
Such methods could strongly increase the precision of estimates of the percolation thresholds in anisotropic models.

In two dimensions, the percolation of grain or void phase are complementary phenomena, that is, 
there is (almost surely) a percolating cluster of the void phase at occupied area fractions below the threshold, but it vanishes (almost surely) above the threshold.
We can therefore directly apply the numerical results and the approximations also to void percolation and the critical porosity.

The Minkowski functionals help to study the relation between geometry, \eg, quantified by area or perimeter, and topology, \eg, quantified by the Euler characteristic.
They thus allow for insights into percolation as a topological phase transition.
Their generalization, the mixed volumes, provide an explicit formula for the well-known excluded area approximation of percolation thresholds, see (\ref{eq:A_ex})--(\ref{eq_perco_Aex_approx}).
If the percolation thresholds of squares with different orientation distributions are used as empirical parameters, we find that the excluded area
approximation is in relatively good agreement with the percolation thresholds of the rectangles, see Fig.~\ref{fig:excluded_area_approximation}.
For the systems studied here, it provides a close upper bound, which implies that the critical average number of bonds per grain slightly decreases for the rectangles compared to squares with the same orientation distribution.

In contrast to the excluded area approximation, the zero of the Euler characteristic serves as a purely geometric approximation of the percolation threshold.
We show that it provides a lower bound that captures well the qualitative dependence on the system parameters, see figure~\ref{fig_perco_Euler_zero}.
If an empirical parameter is introduced (similar to the excluded area approximation), it allows for quite precise predictions of $\gamma_c$ within a narrow range of parameters studied here, see figure~\ref{fig:offset_EulerZero}.
 
Using the second moments of the Minkowski functionals, we suggest and compare new candidates for accurate approximations of the percolation thresholds, see Figs.~\ref{fig_perco_var} and \ref{fig_perco_cov}.
Some are found to be in surprisingly good agreement for the four systems that are tested here, for example, the minimum point of the covariance of area and Euler characteristic.
More examples and further research are needed to test this suggestion.
An important prospect for applications is that no empirical parameters would be needed because analytic formulas are known for the second moments of the Minkowski functionals~\cite{HugLastSchulte2013, hug_second_2016}.
Similar techniques could also easily be applied to percolation on lattices for which the covariances of the Minkowski functionals can be calculated straightforwardly~\cite{goring_morphometric_2013, Klatt2016}.
Moreover, our results can also lead to more fundamental questions.
Why is there such a good agreement between the percolation threshold and some of the extremum points of the second moments of the Minkowski functionals as a function of the intensity?
Is it a coincidence or is there a connection between local fluctuations and the global topology?

\ack

We thank the German Research Foundation (DFG) for the Grants No. HU1874/3-2, No. LA965/6-2, and No. ME1361/11 awarded as part of the DFG-Forschergruppe FOR 1548 ``Geometry and Physics of Spatial Random Systems.''

\appendix

\section{Simultaneous extrapolation of vertical and horizontal effective percolation thresholds}
\label{sec:appendix}

Based on the finite size scaling in (\ref{eq:scaling_xy_equiv}), we determine the isotropic percolation threshold $\phi_c$ via a simultaneous extrapolation of the effective percolation thresholds in $x$- and in $y$-direction.
More precisely, we estimate the percolation threshold via a maximum likelihood fit of the parameters $\phi_c$, $m^x$, and $m^y$.

The effective percolation thresholds in $x$- and in $y$-directions could be correlated, but the correlation cannot be considered in this analysis.
We neglect the correlation of $\phi_{c,i}^{\mathrm{eff},x}$ and $\phi_{c,i}^{\mathrm{eff},y}$ in the simulation and use the likelihood function
\begin{eqnarray}
  \mathcal{L}=\prod_{L_i}\mathrm{Prob}\left[\phi_{c,i}^{\mathrm{eff},x}|\phi_c,m^x\right]\cdot\mathrm{Prob}\left[\phi_{c,i}^{\mathrm{eff},y}|\phi_c,m^y\right]
\end{eqnarray}
with
\begin{eqnarray}
  \mathrm{Prob}\left[\phi_{c,i}^{\mathrm{eff},z}|\phi_c,m^z\right]=\frac{1}{\sqrt{2\pi}\sigma_{z,i}}\exp\left[{-\frac{\left(\phi_{c,i}^{\mathrm{eff},z}-\phi_c-m^z\cdot
  L_i^{-1/\nu}\right)^2}{2\sigma_{z,i}^2}}\right]\;,
\end{eqnarray}
where for each system size $L_i$ the error of the effective percolation thresholds $\phi_{c,i}^{\mathrm{eff},x}$ and $\phi_{c,i}^{\mathrm{eff},y}$ are given by $\sigma_{x,i}$ and $\sigma_{y,i}$, respectively.
The log-likelihood function can be written as
\begin{eqnarray}
  \log\mathcal{L}=\mathrm{const.} - \frac{\chi_x^2}{2} -
  \frac{\chi_y^2}{2}
\end{eqnarray}
with
\begin{eqnarray}
  \chi_x^2+\chi_y^2=\left(\left(
    \begin{array}{c}
      \phi_{c}^{\mathrm{eff},x}\\
      \phi_{c}^{\mathrm{eff},y}
    \end{array}\right)-F(L^{-\frac{1}{\nu}})\right)^t\mathbf{Cov}^{-1}\left(\left(
    \begin{array}{c}
      \phi_{c}^{\mathrm{eff},x}\\
      \phi_{c}^{\mathrm{eff},y}
    \end{array}\right)-F(L^{-\frac{1}{\nu}})\right)
\end{eqnarray}
and
\begin{eqnarray}
  F(L^{-1/\nu})=\left(
    \begin{array}{c c c}
      {1} & L^{-1/\nu} & 0\\
      {1} & 0 & L^{-1/\nu}
    \end{array}\right)\cdot\left(
    \begin{array}{c}
      \phi_{c}\\
      m^x\\
      m^y
    \end{array}\right)\;,
  \label{eq:percolation_fit}
\end{eqnarray}
where the covariance matrix $\mathbf{Cov}$ is assumed to be diagonal
\begin{eqnarray}
  \mathbf{Cov}=\mathrm{diag}(\sigma_{x,1}^2,\ldots,\sigma_{x,i}^2,\ldots,\sigma_{y,1}^2,\ldots\sigma_{y,i}^2,\ldots)\;.
\end{eqnarray}
Therefore, the maximum likelihood estimation equals a minimum least square fit with block matrices.

\providecommand{\newblock}{}

\end{document}